\documentclass[structabstract]{aa}  
\usepackage{graphicx}
\usepackage{txfonts}
\begin{document}
\title{The VLT-FLAMES Tarantula survey XX. The nature of the X-ray bright emission line star 
VFTS\,399 \thanks{Based on observations collected at the European Southern Observatory under program
 ID 182.D-0222}}

\author{J.~S.~Clark\inst{1}
\and E.~S.~Bartlett\inst{2}
\and P.~S.~Broos\inst{3}
\and L.~K.~Townsley\inst{3}
\and W.~D.~Taylor\inst{4}
\and N.~R.~Walborn\inst{5}
\and A.~J.~Bird\inst{6}
\and H.~Sana\inst{5}
\and S.~E.~de Mink\inst{7,8,9}
\and P.~L.~Dufton\inst{10}
\and C.~J.~Evans\inst{2}
\and N.~Langer\inst{11}
\and J.~Ma\'{i}z Apell\'{a}niz\inst{12}
\and F.~R.~N.~Schneider\inst{13}
\and I.~Soszy\'{n}ski\inst{14}
}
\institute{
$^1$Department of Physics and Astronomy, The Open 
University, Walton Hall, Milton Keynes, MK7 6AA, United Kingdom\\
$^2$Astrophysics, Cosmology and Gravity Centre, Department of Astronomy, University of Cape Town, Rondenbosch 7701, Republic of South 
Africa\\
$^3$Department of Astronomy \& Astrophysics, 525 Davey Laboratory, Pennsylvania State University, University Park, PA 16802, USA\\
$^4$UK Astronomy Technology Centre, Royal Observatory Edinburgh Blackford Hill, Edinburgh, EH9 3HJ, UK\\
$^5$Space Telescope Science Institute, 3700 San Martin Drive, Baltimore, MD, 21218, USA\\
$^6$School of Physics \& Astronomy, University of Southampton, Highfield, Southampton, S017 1BJ, UK \\
$^7$Astronomical Institute Anton Pannekoek, University of Amsterdam, 1098 XH Amsterdam, The Netherlands \\
$^8$Carnegie Institution for Science: The Observatories, 813 Santa Barbara St, Pasadena, CA 91101, USA \\
$^9$TAPIR institute, California Institute of Technology, Pasadena, CA 91125, USA\\ 
$^{10}$Astrophysics Research Centre, School of Mathematics and Physics, Queen's University Belfast, Belfast BT7 1NN, UK\\
$^{11}$Argelander Institut f\"{u}r Astronomie, Auf den H\"{u}gel 71, Bonn, 
53121, Germany\\
$^{12}$Departamento de Astrof\'{i}sica, Centro de Astrobiolog\'{i}a (INTA-CSIC), 
Campus ESA, Apartado Postal 78, 28\,691 Villanueva de la Ca\~{n}ada, 
Madrid, Spain\\
$^{13}$Department of Physics, Denys Wilkinson Building, Keble Road, Oxford, OX1 3RH, United Kingdom\\
$^{14}$Warsaw University Observatory, A1. Ujazdowskie 4, 00-478, Warszawa, Poland
}

   \abstract{The stellar population of the 30 Doradus star-forming region 
in the  Large Magellanic Cloud contains a subset of apparently single, 
rapidly rotating O-type stars. The physical processes leading to the 
formation of this cohort are currently uncertain.}
{One member of this group, the late O-type star VFTS\,399, is found to be 
unexpectedly X-ray bright for its bolometric luminosity - in this study we aim to determine its physical nature and the cause of this behaviour.}
{To accomplish this we performed a  time-resolved analysis of optical, infrared and X-ray observations.}
{We found VFTS\,399  to be an aperiodic photometric  variable with an apparent near-IR excess. Its optical spectrum demonstrates complex emission  
profiles in the lower Balmer series and select He\,{\sc i} lines - taken together these suggest an OeBe classification. The highly variable X-ray  
 luminosity is too great  to be produced by a single star, while the hard, non-thermal nature suggests the presence of  an  accreting relativistic 
 companion.  Finally, the detection of periodic modulation of the X-ray lightcurve is most naturally explained under the assumption that the 
accretor is a neutron star.}
{VFTS\,399 appears to be  the first high-mass X-ray binary identified within 30 Dor, sharing many observational  characteristics with classical Be X-ray binaries. 
 Comparison of the current properties of VFTS\,399 to binary-evolution models suggests a progenitor mass  $\gtrsim25M_{\odot}$ for the putative 
neutron star, which may host a magnetic field comparable in strength to those of magnetars.  VFTS\,399  is now the second member of the cohort of 
rapidly rotating  `single' O-type stars in 30 Dor to show evidence of binary interaction 
resulting in spin-up, suggesting that this may be a viable evolutionary pathway for the  formation of a subset of  this stellar population.}


\maketitle

\section{Introduction}

Located within the Large Magellanic Cloud (LMC), the 30 Doradus complex is the most luminous H\,{\sc ii} region in the Local Group. It contains $>10^3$ OB stars, with star formation apparently commencing $\sim$25Myr ago and continuing to the present day  (Walborn \& Blades \cite{walborn97}, Grebel \& Chu \cite{grebel}, Walborn et al. \cite{walborn13}). 
Combined with a moderate line-of-sight extinction, 30 Dor therefore provides an exceptional opportunity to investigate the complete life-cycle of massive stars. In order to exploit this potential, a multi-epoch  spectroscopic observing campaign - the VLT-FLAMES Tarantula Survey (VFTS) - was undertaken between 2008 October-2009 February; a full description of this investigation, including target list, observational strategy and data-reduction methods employed was presented by Evans et al. (\cite{evans}).

 Fulfilling a design goal of the programme, the multi-epoch observations enabled the identification of  both single and binary OB 
star populations  (Sana et al. \cite{sana13}).  Investigation of the single O-type stars revealed that the  
distribution of their  rotational velocities has  a two-component 
structure (Ram\'{i}rez-Agudelo et al. \cite{oscar}), comprising a low velocity peak ($\sim80$kms$^{-1}$) and a high-velocity tail 
(extending to $\sim600$kms$^{-1}$).
The origin of the high velocity cohort is currently uncertain, with the relative contribution from  natal and binary-driven channels still ill-constrained (de Mink et al. \cite{demink}, Ram\'{i}rez-Agudelo et al. \cite{oscar}).  With a projected equatorial rotational velocity, $v_e$sin$i \sim334\pm18$kms$^{-1}$ (Ram\'{i}rez-Agudelo et al. \cite{oscar}), the O9 IIIn star VFTS\,399 is a member of this group (Walborn et al.  \cite{walborn}), while its location on the periphery of the  30 Dor complex (Fig. 1) is suggestive of   a potential runaway nature (cf. Blaauw \cite{blaauw}).  Unexpectedly, cross correlation of the VFTS sample with the X-ray survey of Townsley et al. (\cite{townsley14}) revealed it to be both highly variable and, at its peak, one of the most X-ray luminous stars within 30 Dor. 
Motivated by these observational findings we report the results of a multi-wavelength  analysis of VFTS\,399  in order to clarify its nature, presenting the data in Sect. 2, utilising these to infer its physical properties in Sect. 3, discuss the implications of these findings in Sect. 4, before summarising our results in Sect. 5.

\begin{figure}
\includegraphics[width=8.5cm,angle=0]{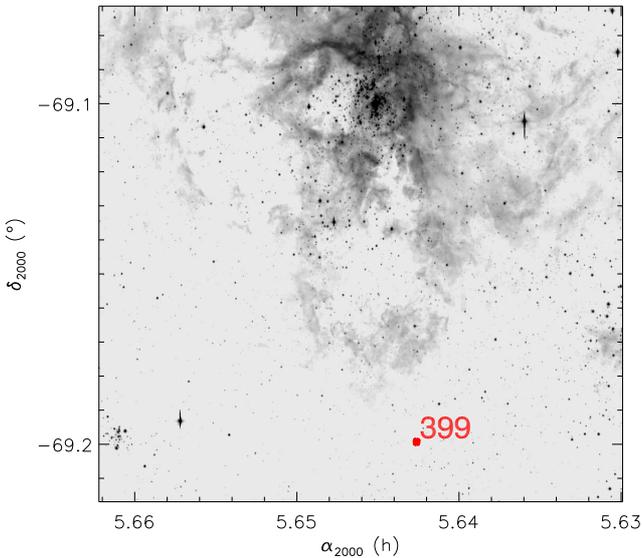}
\caption{WFI image of the southern reaches of 30 Dor with  VFTS\,399  indicated, 
highlighting its location on the periphery of the complex. The field of view is $\sim135pc{\times}125pc$ for a LMC distance modulus of 18.5mag (e.g. Gibson \cite{gibson}).}
\end{figure}

\section{The observational properties of VFTS\,399 }

\subsection{Optical and IR properties}

\subsubsection{Photometry}

Evans et al. (\cite{evans}) provided single-epoch optical and near-IR photometry of VFTS\,399, which we present in Table 1. The latter
are derived from the survey of the LMC by Kato et al. (\cite{kato}), which were obtained on 2003 November 16. Earlier 2MASS  magnitudes, 
obtained on 1998 March 19, are also available (Skrutskie et al. \cite{skrutskie})\footnote{$J=15.48$, $H=15.61$ and $K_{\rm s}=14.83$, noting the $K_{\rm s}$-band magnitude is flagged as uncertain.} and  are suggestive of variability (see below). Finally {\em Spitzer} SAGE observations 
(Meixner et al. \cite{meixner}) obtained between 2005 July 15-26 and October 26-November 2  are also presented in Table 1.

 VFTS\,399  also lies within fields covered by the Optical Gravitational Lensing Experiment (OGLE-III; Udalski 
\cite{udalski}) and VISTA-VMC (Cioni et al. \cite{cioni}) surveys. Unfortunately, neither dataset
coincides with the period over which spectroscopic observations were made but both clearly indicate that 
VFTS\,399  is variable at the level of $\sim0.1-0.2$mag over timescales of $\sim10-10^2$~days in the $V, I$ and $K_{\rm s}$-bands (Fig. 2).
In the extensive $I-$band dataset the variability appears characterised by significant scatter superimposed on 
longer-term variability. The
$K_{\rm s}$-band data are more sparsely sampled, but are suggestive of coherent variability over a $\sim10^2$ day 
 timescale. Preliminary analyses of the more extensive OGLE-III $I-$band dataset revealed a number of potential periodicities  - e.g. $\sim156, 370$ and 720~days - although the 
later  two appeared  to result from the seasonality of the observations. Neveretheless, in order to 
 assess the significance of all the periodicities we followed the methodology employed by Bird et al. (\cite{bird})\footnote{A fast implementation of the Lomb-Scargle 
periodogram, 
with periods 
 searched for ranging from 2 days (the Nyquist frequency) to $\sim$1000 days (dictated by the length of the lightcurve). A rolling-mean was further calculated for the lightcurve  
 and was used to de-trend the data, with extensive Monte-Carlo simulations employed to determine the significance and associated errors of any periods detected. See Bird et al. 
 (\cite{bird}) for further details.}. No coherent statistically-significant periods were identified for the complete dataset, while searches over subsets of these data also 
returned no 
 convincing evidence for transient periodicity.

\begin{table}
  \caption[]{Optical and IR photometry of VFTS\,399. } 
\begin{center} 
\begin{tabular}{ccccccc}
\hline\hline 
$B$ & $V$ & $J$ & $H$ & $K_{\rm s}$ & [3.6] & [4.5] \\
\hline
15.91 & 15.83 & $15.40 $ & $15.30$ & $15.22$ & $14.95$ & $14.95$\\
\hline
\end{tabular} 
\end{center} 
\end{table}

\begin{figure}
  \begin{center}
       \includegraphics[width=8.5cm]{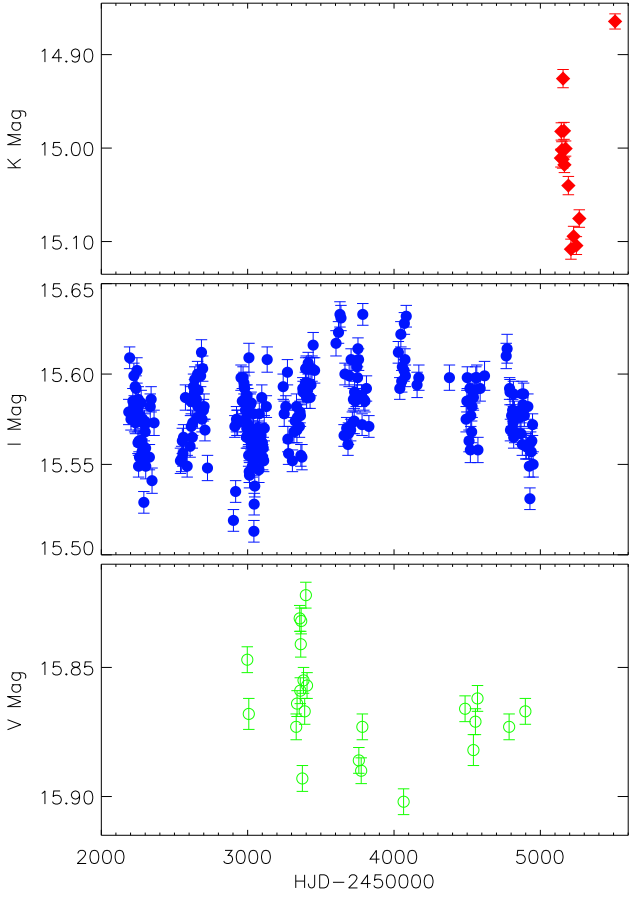} 
  \end{center}
  \caption{VISTA-VMC $K_{\rm s}$-band (upper  panel) and  OGLE $I-$(middle panel) and $V-$band (lower panel) lightcurves for VFTS\,399 .}
\end{figure}

\subsubsection{Spectroscopy}
The VLT/FLAMES spectroscopy of VFTS\,399  is summarised in Table 2 and the 
data-reduction 
techniques employed are fully described in Evans et al. (\cite{evans}), noting that sky subtraction was accomplished
utilising a median sky spectrum constructed from all the sky fibres of the frame that were uncontaminated by flux from bright stars.
 Walborn et al. (\cite{walborn}; their Fig. 10)  presented the 3960-4750{\AA} portion of the spectrum from which they assigned an O9 IIIn
classification; for brevity we do not replicate this here,
 but do plot the H$\alpha$, H$\beta$ and He\,{\sc i} 6678{\AA} emission line profiles in Fig. 3.

 All three lines demonstrate complex, morphologically similar profiles with 
pronounced high-velocity  emission,  of comparable intensity in both blue and red line wings and a strong, narrow component 
centred on  the systemic velocity of VFTS\,399.  Similar, single-peaked emission at low projected velocities 
 is  also present in the photospheric profiles of the higher Balmer series and select He\,{\sc i} lines (Walborn et al. 
\cite{walborn}). Given the presence of spatially-variable nebular emission across 30 Dor we caution against the  over interpretation of the 
profiles of the 
core components in these lines. Sky subtraction successfully removed
 the  [N\,{\sc ii}] 6547{\AA} nebular line, but left small single-peaked residuals in the [N\,{\sc ii}] 6584{\AA} and [S\,{\sc ii}] 6717+6730{\AA} lines. Consequently we suspect that 
nebular contamination might also be present at low projected velocities in the H$\alpha$, H$\beta$ and He\,{\sc i} 6678{\AA} transitions. 
We discuss the physical interpretation of these profiles further in Sect. 3.

Ram\'{i}rez-Agudelo et al. (\cite{oscar}) determined $v_e$sin$i \sim334\pm18$kms$^{-1}$   from the 
LR02 and LR03 spectra, while Sana et al. (\cite{sana13}) utilised  the multi-epoch nature of the  dataset to search for 
radial velocity (RV) shifts. They found that VFTS\,399  demonstrated
epoch-to-epoch RV shifts with a peak-to-peak amplitude of $\sim50$kms$^{-1}$. Given the magnitudes of the observational uncertainties (Table 2), this
is significant at the $\sim2.7\sigma$ level, meaning that VFTS\,399   fell  below the conservative threshold set by these authors
for classification as a binary. We return to this issue in Sect. 3.1, noting that  these observations also yield a  mean RV value of  
$289\pm19$kms$^{-1}$.

\begin{table}
  \caption[]{Summary of spectroscopic observations.} 
\begin{center} 
\begin{tabular}{lcccc}
\hline\hline 
Mode +  & $\lambda$-coverage & $R$ & Date & RV \\
Setting &  ({\AA})           &     &  & (km s$^{-1}$)\\
\hline
Medusa LR02  & 3960-4564 & 7000  & 22/12/08 & $266.8\pm13.3$ \\
             &           &       & 24/12/08 & $286.5\pm9.1\phantom{1}$ \\
             &           &       & 27/01/09 & $300.3\pm17.3$\\
             &           &       & 17/02/09 & $265.8\pm16.0$\\
             &           &       & 07/10/09 & $276.5\pm19.2$\\
Medusa LR03  & 4499-5071 & 8500  & 20/12/08 & $304.9\pm10.6$\\
             &           &       & 21/12/08  & $317.1\pm12.7$\\
             &           &       & 22/12/08 & $297.3\pm13.4$\\
Medusa HR15N  & 6442-6817 & 16000 &  18/12/08 & - \\
              &           &       &  19/12/08 & -\\
\hline
\end{tabular} 
\end{center}
{The He\,{\sc ii} 4200{\AA},   He\,{\sc i} 4387{\AA} and  He\,{\sc ii} 4541{\AA} transitions were employed for $RV$ determination with Medusa LR02 observations, and He\,{\sc i} 4713{\AA} and He\,{\sc i} 4922{\AA} with Medusa LR03.
Note the LR02 and LR03 observations made on 2008 Dec. 22 were separated by $\sim68$mins.} 
\end{table}

\begin{figure}
\includegraphics[width=8.5cm,angle=0]{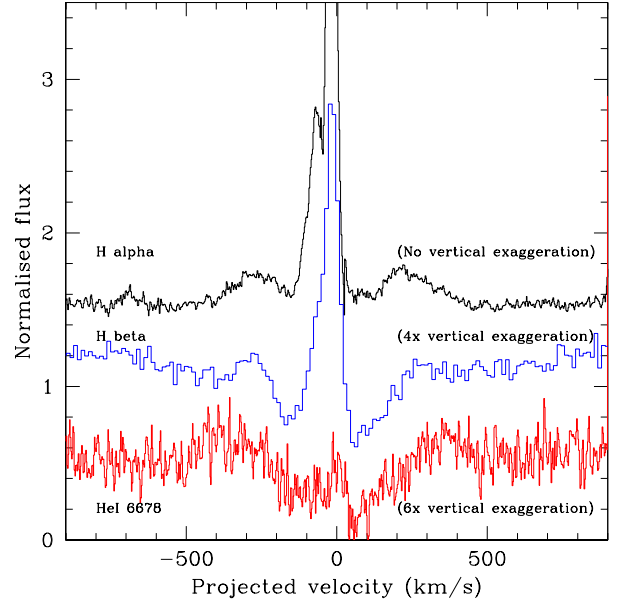}
\caption{ Montage of the H$\alpha$, $\beta$ and He\,{\sc i} 6678{\AA} line profiles, with 
the latter two plotted with an arbitrary vertical exaggeration and offset in order to ease
comparison to H$\alpha$. Note the central component of H$\alpha$ at projected velocity $\sim0$km s$^{-1}$ has a maximum peak intensity of $\sim6\times$
continuum; we suspect nebular contamination of this feature and the comparable peaks in H$\beta$ and He\,{\sc i} 6678{\AA}.}
\end{figure}

\subsection{X-ray properties}

Fortuitously, VFTS\,399 has been the subject of a number of X-ray observations spanning $\sim15$years.
Both XMM-Newton (XMMU J053833.9-691157; Shtykovskiy \& Gilfanov \cite{shtyk}) and {\em Chandra} (CXOU J053833.4-691158 = 
 [TBF2006] 27; Townsley et al. \cite{townsley06}) provide X-ray point 
source 
detections that may be associated with VFTS\,399. Utilising data from 2001 
November 19, the 
 former study  quotes $L_{\rm  2-10keV}\sim1.1\times10^{35}$erg s$^{-1}$, assuming a power law spectrum with photon index, $\Gamma=1.7$ and $N_{\rm 
H}=6\times10^{20}$cm$^{-2}$. The latter study utilised a $\sim21$ks observation from 1999 September and returned $L_{\rm 0.5-8keV}\sim 10^{34}$erg s$^{-1}$ following a power law fit with $\Gamma=1.7$ and $N_{\rm H}=1.3\times10^{22}$cm$^{-2}$.

Subsequently, further {\em Chandra} observations were made on 2006 January 21, 22 and 30 
(Townsley et al. \cite{townsley14}). VFTS\,399 is also located within the target field of the current $\sim2$Ms T-ReX {\em 
Chandra} programme (PI: L.~Townsley) which runs through 2015; consequently we  appraised the 
multiple additional observations available to us at the time of writing (Table 3), deferring an analysis  of the full data set for a future paper.

VFTS\,399  was detected in each observation  and was found to be well separated from other X-ray sources. The resultant 
lightcurve is presented in Fig. 4 and it is immediately apparent that the source is highly 
variable, even within a single observational segment.  In order to screen for variability in the gross morphological properties of the
spectra we  determined the median event energy for individual observations (Fig. 4). This metric shows 
no  evidence for spectral variability across this observational sequence. In order to consolidate  this finding 
we analysed the first eight epochs of data which, when taken together, appear representative of the whole dataset inasmuch as they fully sample the 
full dynamical range found for VFTS\,399 ($L_{\rm 0.5-8keV}\sim 10^{34}-10^{35}$erg s$^{-1}$). Furthermore, excluding the 1999 data due to calibration issues, we also summed all remaining observations to produce a single, high S/N spectrum which we treated in the same manner. 

Specifically, we first subtracted a  background spectrum (corresponding to $\sim3$\% of the extracted 0.5-8.0keV X-ray counts) before several different models within {\tt xspec} (Arnaud \cite{arnaud}) were employed in an attempt to  fit these data. A simple absorbed power-law  ({\tt tbvarabs$^*$pow})
with $Z=0.4\times Z_{\odot}$ was found to provide the best fit (Fig. 5): parameters returned for this model upon application to the summed spectrum were 
$\Gamma=1.0^{+0.1}_{-0.1}$, $N_{\rm  H}=2.7^{+0.4}_{-0.3}\times10^{22}$cm$^{-2}$ (corresponding to $A_{\rm V}\sim16.9^{+2.6}_{-1.9}$mag, suggestive of significant obscuration intrinsic to VFTS\,399) and an absorption corrected $L_{\rm 0.5-8keV}\sim5.0\times10^{34}$erg s$^{-1}$.

The results obtained from analysis of the first eight individual epochs are presented in Table 4; this yielded  
a range of parameters spanning $0.8 \leq \Gamma \leq 1.1$, $2.4\times10^{22} \leq N_{\rm 
  H} \leq 3.1\times10^{22}$cm$^{-2}$ and $ 1.0\times10^{34} \leq L_{\rm 
0.5-8keV} \leq 12.6\times10^{34}$erg s$^{-1}$. 
 For epochs with very few counts $N_{\rm H}$ and $\Gamma$ were held at the 
 value determined from the summed spectrum, with only the normalisation 
allowed to vary. We caution that  fits are provided for epochs 
  as a whole and hence do not account for the variable behaviour of the 
source over the course of a given observation (cf. Fig. 4). Moreover, 
while some datasets are  fitted by 
this simple model, others are not; this likely indicates deficiencies in the model, but the low count rate precludes more sophisticated treatments at this stage.  Nevertheless, we conclude that there is no evidence for evolution of the gross morphological properties of the X-ray spectrum of VFTS\,399 over the course of the observations.

\begin{table}
\caption{Summary of {\em Chandra} X-ray observations of VFTS\,399 during 2006 and 2014.}
\begin{center}
\begin{tabular}{ccc}
\hline
\hline
Obs. ID & Start Time & Exposure (s)\\

\hline
 5906 &  2006-01-21T19:04 &   12317 \\
 7263 &  2006-01-22T16:51 &   42528 \\
 7264 &  2006-01-30T15:06 &   37593 \\
16192 &  2014-05-03T04:10 &   93760 \\
16193 &  2014-05-08T10:15 &   75993 \\
16612 &  2014-05-11T02:15 &   22672 \\
16194 &  2014-05-12T20:00 &   31335 \\
16615 &  2014-05-15T08:24 &   45170 \\
16195 &  2014-05-24T14:09 &   44405 \\
16196 &  2014-05-30T00:05 &   67109 \\
16617 &  2014-05-31T01:27 &   58860 \\
16616 &  2014-06-03T22:26 &   34529 \\
16197 &  2014-06-06T12:32 &   67802 \\
16198 &  2014-06-11T20:20 &   39465 \\
16621 &  2014-06-14T14:46 &   44399 \\
16200 &  2014-06-26T20:01 &   27360 \\
16201 &  2014-07-21T22:13 &   58393 \\
16640 &  2014-07-24T11:21 &   61679 \\
16202 &  2014-08-19T15:30 &   65128 \\
17312 &  2014-08-22T06:21 &   44895 \\
\hline
\end{tabular}
\end{center}
\end{table}

\begin{figure*}
  \begin{center}
       \includegraphics[width=18cm, angle=-00]{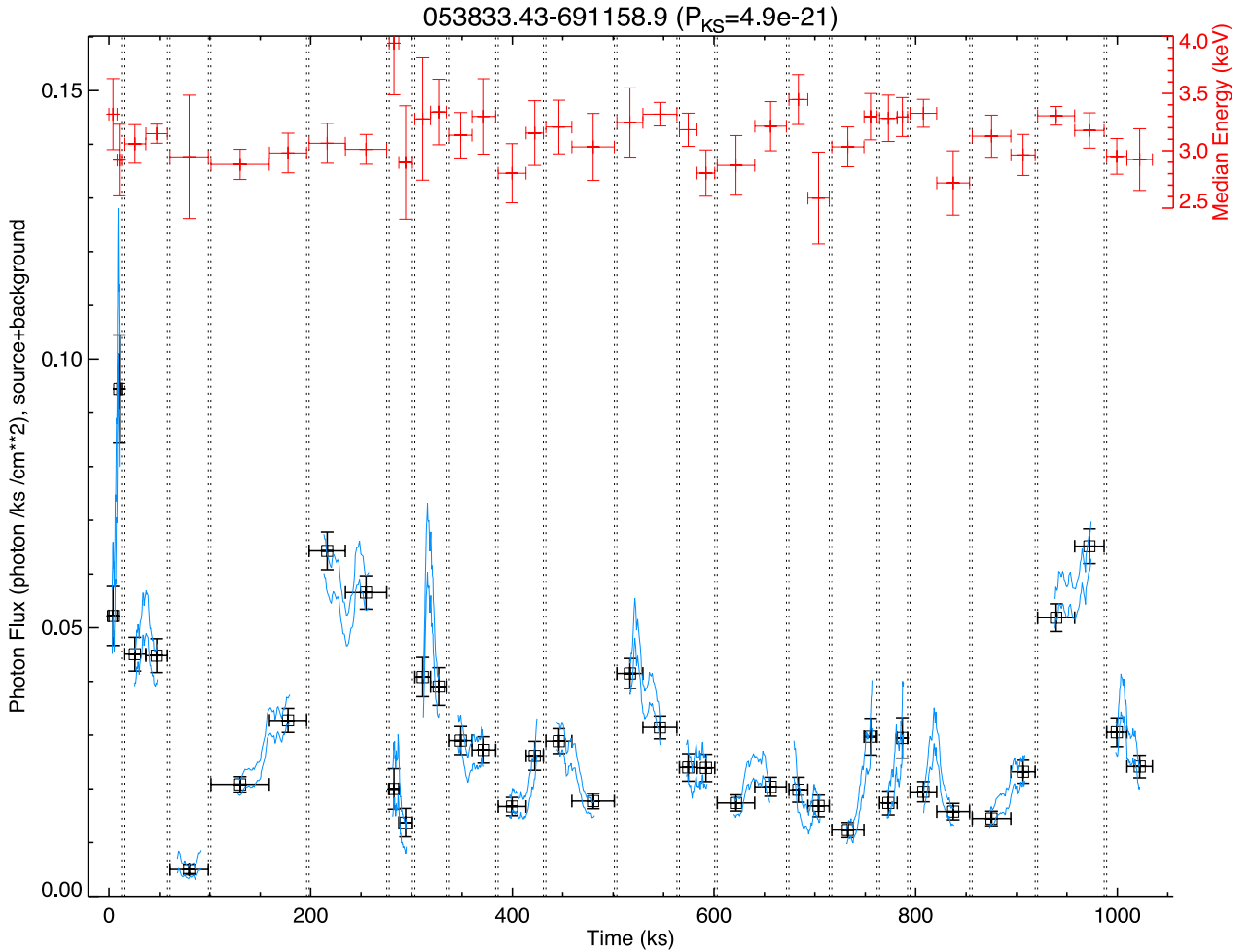} 
  \end{center}
  \caption{Concatenated {\em Chandra} observations of VFTS\,399 (975~ks combined exposure). Binned flux and associated $1\sigma$ errors are shown in black, while binned median energies are shown in red. A sliding-window lightcurve inferred from the individual events (i.e. no binning) is shown by the blue lines, with a 68\% pointwise confidence band limit represented by the two lines. Running left to right, the start dates of the 20 epochs of observations are given in Table 3.}
\end{figure*}

\begin{table*}
\caption{Summary of indicative fits to the {\em Chandra} X-ray observations of 2006 and 2014.}
\begin{center}
\begin{tabular}{ccccccccc}
\hline
\hline
ObsID   &  Date    & Exposure & Net     &  $N_{\rm H}$                &  $\Gamma$           & Normalisation $(\times10^{-5}$      & $\chi^2$/DoF &   log$L_{\rm 0.5-8kev}$ \\
        &          &  (s)     &Counts   & ($\times10^{22}$ cm$^{-2}$) &                     & photons cm$^{-2}$ s$^{-1}$ keV$^{-1})$    &              &  
(ergs$^{-1}$) \\
\hline
Summed  & -        & 975400   & 5397    & $2.7_{-0.3}^{+0.4}$         & $1.0_{-0.1}^{+0.1}$ & $1.5_{-0.2}^{+0.2}$ &   125.5/115    &    34.7 \\

 5906   & 21/01/06 & 12320    & 175     & $2.6_{-1.5}^{+1.9}$         & $0.8_{-0.3}^{+0.5}$ & $2.9_{-1.4}^{+3.0}$ &    4.8/8     &    35.1 \\

 7263   & 22/01/06 & 42528    & 407     & $3.0_{-1.0}^{+1.4}$         & $1.0_{-0.3}^{+0.3}$ & $2.4_{-0.8}^{+1.4}$ &   24.1/23    &   34.9 \\

 7264   & 30/01/06 & 37593    & 40      & {\em 2.7}                   &  {\em 1.0}          & $0.2_{-0.1}^{+0.1}$ &    6.7/9     &   34.0 \\

16192   & 03/05/14 & 93760    & 419     & $3.1_{-1.3}^{+1.8}$         & $1.1_{-0.3}^{+0.4}$ & $1.5_{-0.6}^{+1.2}$ &   29.6/24    &    34.7 \\
16193   & 08/05/14 & 75996    & 657     & $3.0_{-1.0}^{+1.4}$         & $1.1_{-0.3}^{+0.3}$ & $3.3_{-1.1}^{+1.9}$ &   38.6/38    &    35.0 \\

16612   & 11/05/14 & 22678    & 54      & {\em 2.7}                   &  {\em 1.0}          & $0.6_{-0.2}^{+0.2}$ &    9.8/7     &    34.4 \\

16194   & 12/05/14 & 31335    & 238     & {\em 2.7}                   & $1.0_{-0.3}^{+0.3}$ & $1.8_{-0.5}^{+0.6}$ &   30.6/13    &    34.8 \\

16615   & 15/05/14 & 45170    & 237     & $2.4_{-1.3}^{+1.8}$         & $0.8_{-0.4}^{+0.4}$ & $1.2_{-0.5}^{+1.0}$ &    6.4/12    &     34.7 \\
\hline
\end{tabular}
\end{center}
{Summary of the X-ray model parameters for the summed spectrum (top row) and the individual spectra derived from the first eight representative, constituent spectra. Parameters given in italics have been fixed to the properties 
derived from the summed spectrum due to low count rate, note also the comparatively poor  fit to the spectrum of 2014 May 12. In all cases errors
quoted are 90\% confidence limits. The X-ray luminosities are derived after correction for extinction and for a distance to the LMC of $50kpc$.}
\end{table*}

\subsubsection{X-ray timing analysis}

\begin{figure}
  \begin{center}
       \includegraphics[angle=-90,width=8.5cm]{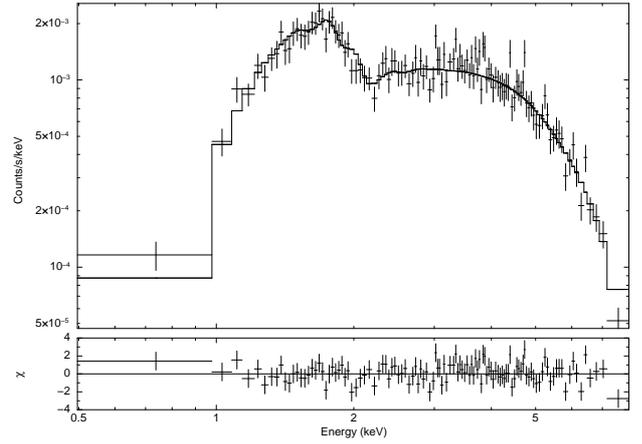} 
  \end{center}
  \caption{Top panel:Absorbed power-law fit to the summed spectrum constructed from the 20 epochs of {\em Chandra} observations shown in Table 3 (5397 net counts; 975ks of combined exposure).  Bottom panel: residuals for this fit.}
\label{Xray_plot}
\end{figure}

\begin{figure}
  \begin{center}
       \includegraphics[angle=-0,width=8.5cm]{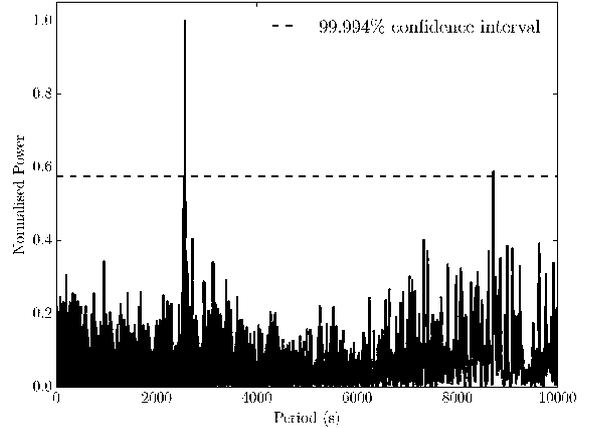} 
  \end{center}
  \caption{LS periodogram for the combined dataset, the dashed line representing the 4$\sigma$ confidence interval, which the peak corresponding to the $\sim2567$~s period clearly exceeds.}
\end{figure}

\begin{figure}
  \begin{center}
       \includegraphics[angle=-0,width=8.5cm]{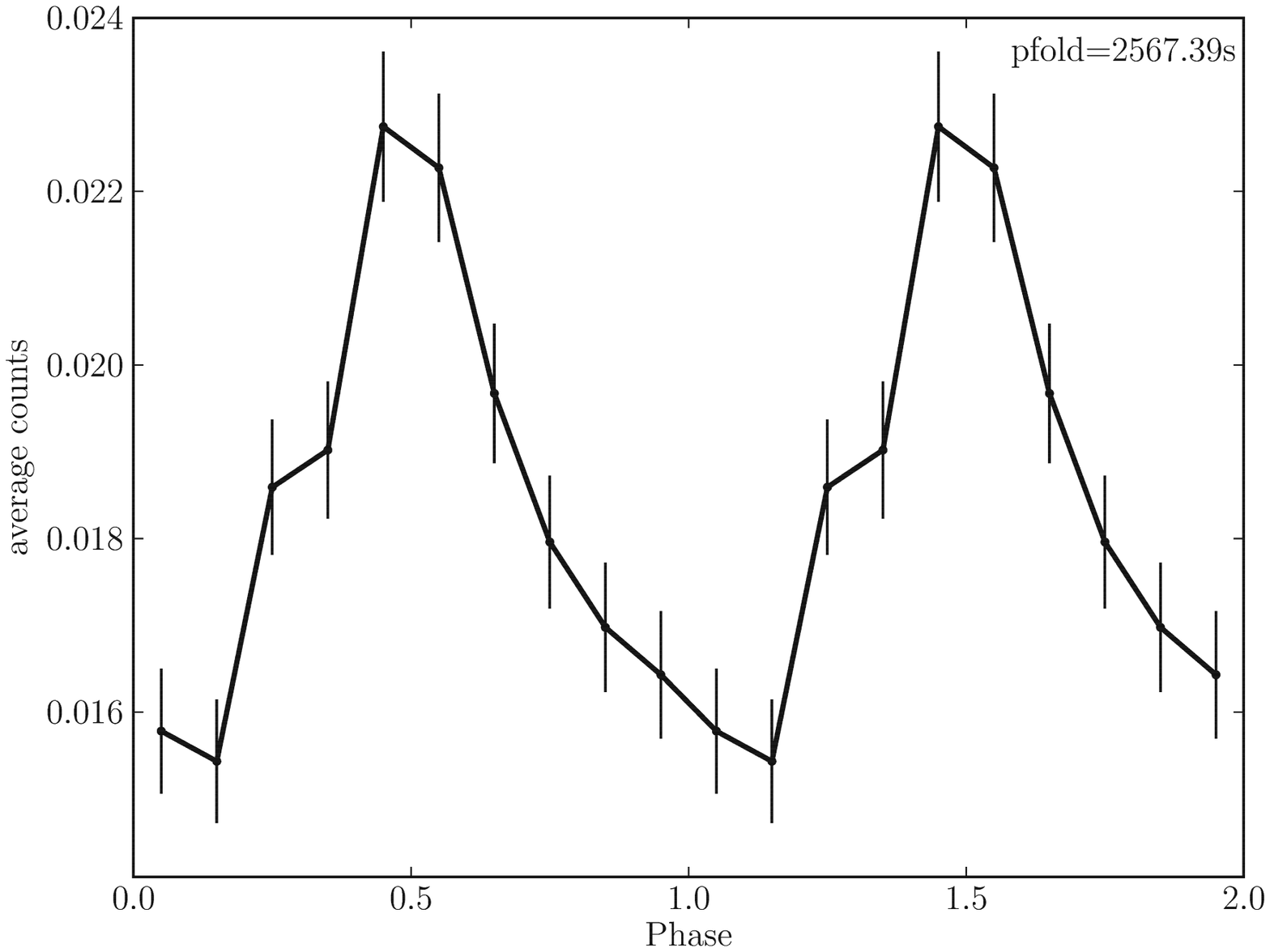} 
  \end{center}
  \caption{X-ray data folded on the $\sim2567$~s period. The pulsed fraction, calculated by  integrating over the profile and computing the ratio between the pulsed section of the profile and the total flux, is $0.17\pm0.04$.}
\end{figure}

 Lomb-Scargle (LS;  Lomb \cite{Lomb76}, Scargle \cite{Scargle82}, Press \& Rybicki \cite{Press89}) analysis was performed on each 
of the single  observation light   curves as well as on a composite light curve, made up of all the data sets. LS analysis is ideal 
for detecting weak, periodic  modulation in unevenly sampled data sets (such as our composite data set). A 'blind' search for periods between 7s (approximately   twice the frame interval of {\em Chandra}) and 1$\times10^5$s was performed with a frequency step of $1\times10^{-7}$. 
Foreshadowing the discussion in Sects. 3.2  and 4.1, this exceeds the observed range of pulsational periods exhibited by neutron stars in X-ray binaries  with OB star primaries.

 At the other extreme, long-term modulation of the lightcurve might be 
 expected from the rotation of magnetic stars, or due to orbital 
 modulation in colliding-wind systems or accreting high-mass 
 X-ray binaries; in  all cases the resultant period may range from  
 several days to several years. Unfortunately, the durations of the 
individual observing blocks are $<1$day (with only a single exception; Table 3) meaning that none would be expected to sample a full cycle of a system with a period of a few days. Moreover, the overall time span of the observations is such that we would also not expect to fully sample a single cycle of any long-period ($\gtrsim10^2$day) modulation. Thus, we conclude that the current dataset is ill-suited to the {\em robust} detection of any such putative period; a longer (continuous) time base of observations would be required.

Furthermore,  orbital modulation of the X-ray light curves of such binaries occurs in the subset of eccentric-orbit systems, which may demonstrate high-amplitude  flaring during perisatron passage due to strengthened wind/wind 
interactions in colliding wind systems (e.g. Corcoran \cite{corcoran}) or enhanced accretion rates in X-ray binaries
(e.g. Galache et al. \cite{galache}). Such repetitive, periodic flaring  
is not 
immediately apparent in the lightcurve of VFTS\,399, although this does not preclude its future occurrence since, in the case of X-ray binaries, such behaviour is observed to be transient.

With the above constraints in mind, Fig.  6 shows the LS periodogram of the whole data set. There appears to be evidence for a period at $\sim2567$~s at $>4\sigma$ level of significance. Fig. 7 shows the pulse profile, which appears asymmetric and single peaked. A peak at this period is present in the periodograms of a number of individual observation blocks, albeit inevitably at a lower level of significance. The strength of the detection  of a period in a given observation depends on the S/N of the observation, the pulsed fraction and the period length: it is easier to find shorter rather than longer periods in any given observation since more 'cycles' are present. As such it is not surprising that this period is not robustly detected in some observations; for instance our shortest observation, 5906, is only 12.3 ks in duration, which allows for \textless5 pulsations.  Nevertheless, this latter result gives us confidence that the period is  not spuriously introduced by our analysis of the  composite  lightcurve during the process of `stitching together' its constituent parts; this assumption was quantitatively tested as described below\footnote{Such a methodology is
 routinely applied in e.g. the case of optical observations of X-ray binaries, where periodic gaps between distinct segments of the lightcurve 
are introduced by seasonal visibility constraints (e.g. Bird et al. \cite{bird}). Moreover, Reig et al. (\cite{reig09}) encountered 
exactly this issue in their search for X-ray pulsations in the high-mass binary 4U2206+54, due to visibility gaps in their {\em RXTE} 
lightcurve.}. Finally, a strong period   at 707~s detected in the periodogram of observation 16193 is the \emph{Chandra} dither period, resulting from   a bad CCD column falling  within the aperture.

In order to test both the robustness and statistical significance of this period a number of  simulations of the data were run. LS analysis was first performed on the barycentric-corrected X-ray event
timestamps with a variety of sizes to make sure that the 2567~s
period was not an artefact 
of the binning adopted  - in each case the period was still detected. The data were then bootstrapped to determine how robust 
the period was. This was accomplished by  creating `fake' data sets by using a random number generator to select   $\sim63$\% of the 
data points and  performing a LS  analysis on the resultant lightcurve. If the detected period was being driven by a small 
number of data points it will not be consistently detected during   the bootstrapping. The period at $\sim2567$~s was detected in 96.6\% 
of the trials, suggesting that the period is indeed robust. 
 
 We also ran a second bootstrapping analysis, 
this time  selecting  random observations (as opposed to data points) to build the simulated data sets. Anywhere from 30 to 80\% 
of the total data  were used (since there are fewer  ways to randomly select 20 light curves in contrast to the full set of data 
points). In this instance 
the period at $\sim2567$~s was recovered in 85\% of the trials. The discrepancy between  the two analyses almost certainly arises 
in the low data fraction regime in the second bootstrapping, since recovering the period with  only $30$\% of the original data will be 
unlikely.

The significance of the detected period was assessed by estimating
the probability of the detection under the null hypothesis that the
lightcurve is uniform. Over
15000 such light curves were generated by randomizing the X-ray
event timestamps.
 This allows us to create simulated light curves with the same statistical properties as  the original data set. LS analysis 
was performed on each of these light curves and the maximum LS power achieved recorded. This  is shown as the dashed line in 
Fig. 6. The results of the Monte-Carlo analysis suggest that our $\sim2567$~s period, with  a
 LS power of 42.8 has a less than 1 in 15787  (i.e. \textless0.006334\%) 
probability of being  a chance occurrence due to white noise. Although a 
second peak (at ~8718 s) in the periodogram exceeds this
significance threshold, it  is not robust to the above bootstrapping 
analyses; specifically, the  peak disappears into the noise if the three 
observing blocks from 2006 are excluded, suggesting that it is most likely 
an artefact.

We therefore conclude that the Monte-Carlo and bootstrapping  analyses  both suggest  there is a periodic 
modulation in our data set although, trivially, it cannot reveal the origin of this periodicity. 
The $\sim2567$~s period does not correspond to the \emph{Chandra} frame interval (3.241040~s), dither  period (707~s) or the orbital period 
(198~ks), nor 
does it correspond to an alias (the beat frequency) of any of these three potential  periods with each other (which corresponds to
3.241093, 3.255966~s and 707.5335~s; for a  comprehensive discussion on aliasing and periods see Bird et al. \cite{bird}). For the remainder of the paper we therefore proceed under the assumption that the period is astrophysical in origin.

Typically, in determining the error on a period one assumes  that the period derivative is zero throughout the observation and hence that 
the dominant source of errors is the Poisson error on the counts themselves\footnote{In such  circumstances one may vary the 
lightcurve within the errors on the individual data points in order to  produce a large number (i.e. $\sim10^5$) of synthetic  lightcurves that 
are statistically identical to the original and subject each to a 
  LS search. A histogram of the resultant distribution of periods is typically well approximated by a Gaussian, from which the width  (i.e. standard deviation) yields a quantitative estimate of the period error.}. However, given the time span of our observations we may not {\em 
a priori}  make this assumption. This in turn prevents a robust quantitative estimate of the error on the period since we may not distinguish between random error and a systematic  evolution of the period during the course of the observation.

However, this turns out to be of secondary concern as it is possible that the $\sim2567$s period we detect could be an alias of the {\em true} astrophysical period and the  \emph{Chandra} frame interval (3.241040~s; note that aliasing is  a common problem in the analysis of e.g. optical light curves of X-ray binaries (Bird et al. \cite{bird})). This would require the true period of VFTS\,399 to be  
$\sim3.24515$~s. Unfortunately following the above discussion such a period  would not be {\em directly} detectable in our current dataset; observations by a different observatory such as {\em XMM} would be necessary to distinguish between these possibilities. It is, however, noteworthy that our simulations indicate the $\sim2567$s period is robust to evolution of the period, such as those commonly seen in X-ray binaries due to accretion torques and doppler motion, but {\em not} so if it is due to an alias. Given  the nature of the system (Sect. 3.2 and 4.1) this might be taken as evidence to disfavour the alias hypothesis.

Nevertheless, it is important to emphasise that either interpretation   still requires a real periodicity in the X-ray lightcurve to be present. However, we are left in the uncomfortable position of having identified a statistically-significant period of apparently astrophysical origin in our dataset but not being able to {\em conclusively} determine whether the 
 $\sim2567$~s periodicity is an alias or not. As a consequence for the remainder of the paper (e.g. Sects. 3.2 and 4) we are forced 
to consider the physical implications of both the `short' and `long' periods in parallel. 
Clearly,  independent observations of VFTS\,399 with  observatories other than {\em Chandra} would be highly desirable, since one requires a much shorter frame time in order to test the alias hypothesis.

\section{The nature of VFTS339}

VFTS\,399  is displaced from the locus occupied by the majority of OB stars in both near-IR ($(J-H)/(H-K_{\rm s})$; Dougherty et al. \cite{dougherty})  and mid-IR ([3.6]/([3.6]$-$[4.5]); Bonanos et al. \cite{bonanos}) colour/colour and colour/magnitude diagrams. It is in a region populated  by, amongst others,  early-type stars exhibiting the classical Be phenomenon. Furthermore, photometric variability, such as that demonstrated by VFTS\,399, is a well-known property of classical OeBe stars (e.g. Mennickent et al. \cite{mennickent}).

 Emission in the Balmer series is another  key classification criterion for classical OeBe stars. Under such a scenario, the 
 emission at large projected velocities  in the line wings of H$\alpha$, H$\beta$ and He\,{\sc i} 6678{\AA} (Fig. 3) would result 
 from the presence of a compact, quasi-Keplerian decretion disc; during the early stages of disc-reformation, time resolved 
 H$\alpha$ spectra of $o$ And revealed emission components with a comparable peak-to-peak separation (Clark et al. 
\cite{clark03}). 
Similar behaviour is also observed  in both the H$\alpha$  and He\,{\sc i} 6678{\AA}  lines of the B0 Ve primary of the high mass 
X-ray binary X Per. Comparison of X Per to VFTS\,399  appears particularly apposite,  given that both stars demonstrate 
additional emission features interior to those present in the line  wings. In X Per  this is attributed to the presence of a 
gaseous circumstellar ring exterior to, and rotating more slowly than,  the inner disc  that is responsible for the high-velocity 
emission component of the line profile (Tarasov \& Roche \cite{tarasov}, Clark et al. \cite{clark01})\footnote{In  this regard we 
are mindful of the likelihood of nebular contamination in VFTS\,399. However even if all emission interior to the high velocity wings were nebular in origin - which appears unlikely upon comparison of  the relative widths of the nebular lines to this component - the global consideration that VFTS\,399  is a classical OeBe star would not be affected. However, potential unresolved nebular contamination does preclude the measurement of the line equivalent widths.}.

The  OeBe phenomenon  extends to stars with spectral types as early as that of  VFTS399 (e.g. Neguerueula et al. \cite{iggy04}), while such stars are also 
known to be rapid rotators (e.g. Townsend et al. \cite{townsend}). In light of these observational properties,  VFTS\,399  appears to be a  classical OeBe star
 and hence we slightly revise its formal classification to O9 IIIne. 

\subsection{Stellar parameters}

The broad-band photometry  presented in Table 1 was  analysed using  the Bayesian code CHORIZOS 
in order to determine the interstellar reddening  (Ma\'{i}z Apell\'{a}niz \cite{maiz04}, Ma\'{i}z Apell\'{a}niz et al. \cite{maiz14}),
from which Walborn et al. (\cite{walborn}) inferred  $M_{v}\sim -4.41$, $T_{\rm eff}\sim32.8$kK and log$(L_{\rm bol}/L_{\odot})\sim4.9\pm0.1$ for VFTS\,399.
As part of a wider programme (Ram\'{i}rez-Agudelo, in prep.), the LR02 and LR03 spectroscopic data were subsequently analysed with the non-local thermodynamic equilibrium code FASTWIND (Puls et al. \cite{puls}), yielding $T_{\rm eff}\sim31.2_{-0.5}^{+1.0}$kK, log$g \sim3.6_{-0.1}^{+0.15}$ and log$(L_{\rm bol}/L_{\odot})\sim4.8\pm0.1$.

We have also attempted to estimate the nitrogen abundance in VFTS\,399. 
Due to its large projected rotational velocity, the N\,{\sc ii} spectrum 
could not be analysed but the N\,{\sc iii} multiplet near 4515\AA\ was 
observed as a blend in the LR02 and LR03 spectra. In both spectra, this 
region lies close to either the upper or lower wavelength limits making 
normalisation of the spectra difficult. We have therefore chosen to treat the spectra separately, with a simple subjective normalisation. These spectra are shown in Fig. 8,  together with predictions from model atmosphere calculations. The latter were taken from the O-type star grid of Lanz \& Hubeny (\cite{lanz}) and have been convolved with an appropriate broadening function. The weaker theoretical N\,{\sc iii} spectrum has been interpolated to the FASTWIND atmospheric parameters (and a microturbulence of 10kms$^{-1}$) and is for a nitrogen abundance of 7.62 dex and a LMC metallicity. The stronger spectrum is for a nitrogen abundance of 7.92 dex but a galactic metallicity; this leads to more line blanketing, a hotter temperature structure and, for example, enhanced He\,{\sc ii} absorption. Inspection of grid models with different effective temperatures implies that this will lead to an overestimation of the strength of the N\,{\sc iii} spectra compared with an LMC metallicity.

Both the LR02 and LR03 spectra imply a nitrogen abundance of approximately 7.8 dex. However we emphasise that this value and the atmospheric parameters from  the FASTWIND 
 analysis must be treated with  caution due to the limitations of classical non-LTE techniques.  Both the H$\alpha$ 
and H$\beta$ profiles  are subject to  infilling, with    some evidence for similar behaviour  in the H$\gamma$ profile, although 
a definitive  conclusion is  complicated by possible nebular contamination. Moreover we have assumed 
that an isolated stellar model atmosphere is appropriate; the possible systematic errors introduced by the complexity of the system e.g the presence of an accreting secondary (see Sect. 3.2) are difficult to quantify but could be significant. Nevertheless, we are confident that our conclusion that VFTS\,399 demonstrates a pronounced nitrogen enhancement is robust.

Likewise, contamination of the broad-band photometry by continuum emission from the circumstellar disc inferred for  VFTS\,399  has the potential to seriously compromise  reddening, and hence luminosity, estimates. This is evident  upon consideration of the long term photometric behaviour of the BeX-ray binary A0535+26. Over a $\sim16$yr period Clark et al. (\cite{clark99}) report $\Delta V \sim0.4$mag and $\Delta E(B-V)\sim0.13$mag as a result of disc variability (in the sense the system became redder as it brightened). As a consequence if we fail to account for the presence of the circumstellar disc, we risk overestimating the bolometric luminosity of VFTS399  because (i) we will not subtract the excess disc continuum emission and (ii) because we will overestimate the interstellar reddening component and hence over correct for this.

Riquelme et al. (\cite{riquelme}) investigated this problem for Be X-ray binaries in the Galaxy and Magellanic Clouds, finding it to be significant for the latter objects, where
contributions from circumstellar and interstellar reddening are comparable. In the absence of contemporaneous spectroscopic and photometric observations  of VFTS\,399 in a 
discless state we may not determine the interstellar reddening. Moreover, nebular contamination of the  
H$\alpha$ line prevents an equivalent width measurement and from this
an estimate of  the circumstellar reddening  via the relationship   defined by Riquelme et al. 
(\cite{riquelme}).  As a consequence we are forced to adopt a `worst-case' approach, correcting for the maximum 
contribution from circumstellar reddening reported by  Riquelme et al. (\cite{riquelme}) of $E^{cs}(B-V)\sim0.17$mag. Utilising this value, the photometry from Table 1 and the intrinsic colours of an O9 III star
(Martins \& Pez \cite{martins06}) we may infer an interstellar reddening. Then employing the relation from  Riquelme et al. (\cite{riquelme}) between total 
 optical extinction and interstellar reddening - $A_{V}^{\rm total} =3.1E^{is}(B-V)-0.3$mag - for Be X-ray binaries and an appropriate
bolometric correction (from Martins et al. \cite{martins}) we determine log$(L_{\rm bol}/L_{\odot})\sim4.4$ for VFTS\,399.

This value appears surprisingly low in comparison to  either log$(L_{\rm bol}/L_{\odot})\sim5.17$ or  4.72 expected for, respectively,  O9 giants and main sequence stars (Martins et al. \cite{martins}). It is difficult to envisage any physical process, such as binary interaction, that would leave a star this under luminous for its spectral classification, suggesting that we have over corrected for circumstellar  contamination. Therefore we favour the  intrinsic luminosity of an O9 V star as an appropriate lower bound for VFTS\,399\footnote{Foreshadowing the discussion in Sect. 4.1, Negueruela \& Reig (\cite{negueruela}) reported that the emission-line mass-donor BD$+53^{\rm o} 2790$ (classified as O9.5 V) in the high-mass X-ray binary 4U 2206+54 demonstrates variable emission, infilling
the He\,{\sc ii} 4686{\AA} photospheric line that forms part of the 
luminosity classification criteria for late O-type stars (Walborn et al. 
\cite{walborn}).  If such an effect is also present here then it could erroneously lead to the assignment of a giant rather than dwarf luminosity class for VFTS\,399.} and  the upper bound from Walborn et al. (\cite{walborn}),
leading to log$(L_{\rm bol}/L_{\odot})\sim4.7 - 4.9$.

We employed the Bayesian code BONNSAI\footnote{The {\textsc BONNSAI} web-service is available at 
http://www.astro.uni-bonn.de/stars/bonnsai} 
(Schneider et al. \cite{schneider}) to determine the current mass ($M_{\rm curr}$) and age of VFTS\,399 (Table 5). 
 Firstly, we neglected the luminosity estimate and simultaneously matched the observed effective temperature, 
 $T_{\rm eff}=31.2^{+1.0}_{-0.5}$kK, surface gravity, log$g=3.60^{+0.15}_{-0.10}$ and projected surface 
 rotational velocity, $v_{\rm e} \sin i=334\pm33$kms$^{-1}$ to the rotating, single star LMC models of Brott et 
al. (\cite{brott}) and K\"{o}hler et al. (\cite{koehler}). We used a Salpeter initial mass function (Salpeter 
\cite{salpeter}) and the observed distribution of rotational velocities of 
the 30~Doradus O-type stars (Ram\'{i}rez-Agudelo et al. \cite{oscar}) as 
appropriate priors, together with a uniform distribution of stellar ages and a random orientation of rotational axes.
The stellar models predicted a luminosity of log$(L_{\rm bol}/L_{\odot})=5.0^{+0.25}_{-0.16}$, in agreement with the luminosity estimate of Walborn et al. (\cite{walborn}) and the expectations from the O9 classification. A luminosity of log$(L_{\rm bol}/L_{\odot}) \sim 4.4$ was excluded at more than $3\sigma$. 

Secondly, we took the luminosity estimates for VFTS\,399 into account. Following from above we assumed log$(L_{\rm bol}/L_{\odot})\sim4.7 - 4.9$ but, for completeness, also undertake calculations assuming both log$(L_{\rm bol}/L_{\odot})\sim4.4$ - appropriate for the maximum expected contamination by a circumstellar disc - and log$(L_{\rm bol}/L_{\odot})\sim5.2$ - the expected luminosity for an O9 III star (Martins et al. \cite{martins}). The results are presented in Table 5. For log$(L_{\rm bol}/L_{\odot})\sim4.4$, the stellar models could not reproduce the observables simultaneously, likely due to the low luminosity. For our favoured range of bolometric luminosities we determined $M_{\rm curr}\sim18.0-20.0M_{\odot}$. 
Finally, we inferred a current age of $\sim6$Myr for VFTS\,399, assuming 
single-star evolution. However, 
pre-empting the discussion in Sect. 3.2, we consider it likely that VFTS\,399 has accreted mass from a 
binary companion in the past, rejuvenating the star (e.g. Schneider et al. \cite{rejuvenation}); therefore we 
consider this value as a lower limit for its true age. 

 Under this assumption it is of interest to determine whether the observed nitrogen enhancement in VFTS\,399 is the  result of  binary mass transfer. Langer (\cite{langer}) 
 showed that for Galactic stars such interaction may result in an enrichment of $\sim0.5 - 0.8$dex. Since the initial C/N ratio is $\sim0.2 - 0.3$dex larger in the Galaxy in 
 comparison to the  LMC we would expect the nitrogen enrichment in the LMC to be larger by this amount; for a baseline nitrogen abundance of $\sim6.9$dex in the LMC we would 
predict a post-interaction abundance of $\sim 7.6 - 8.0$dex. Unfortunately, while this is consistent with our findings for VFTS\,{399}, it is also consistent with the predictions for rotational mixing in a single, rapidly rotating star, with BONNSAI predicting $\sim7.76^{+0.24}_{-0.12} - 7.83^{+0.22}_{-0.15}$ dex for our favoured range of luminosities. Therefore we are current unable to utilise the chemical abundance of VFTS\,399 in isolation  to distinguish between single and binary star evolution.

\begin{figure}
\begin{center}
\includegraphics[angle=-0,width=8.5cm]{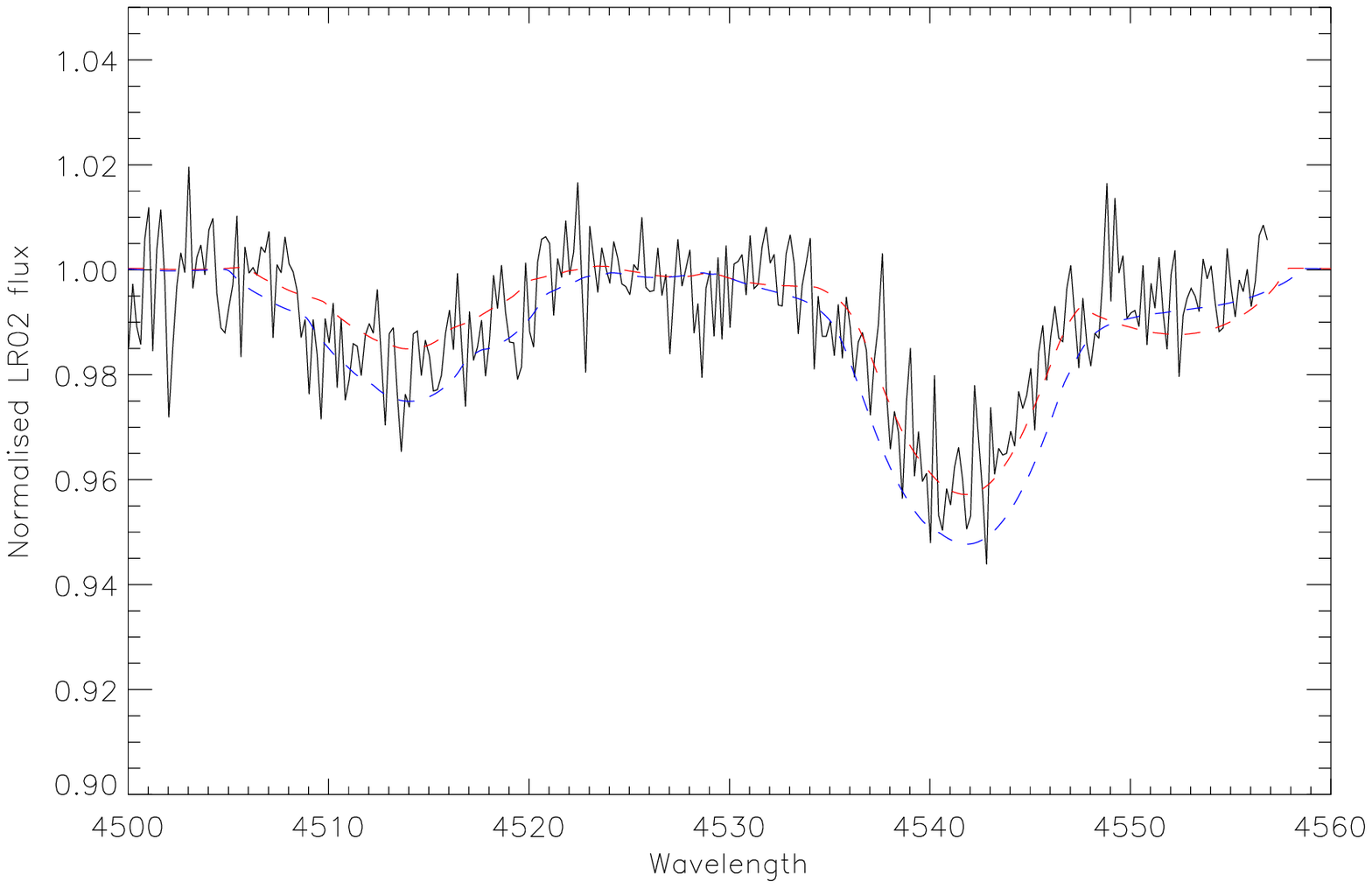}
\includegraphics[angle=-0,width=8.5cm]{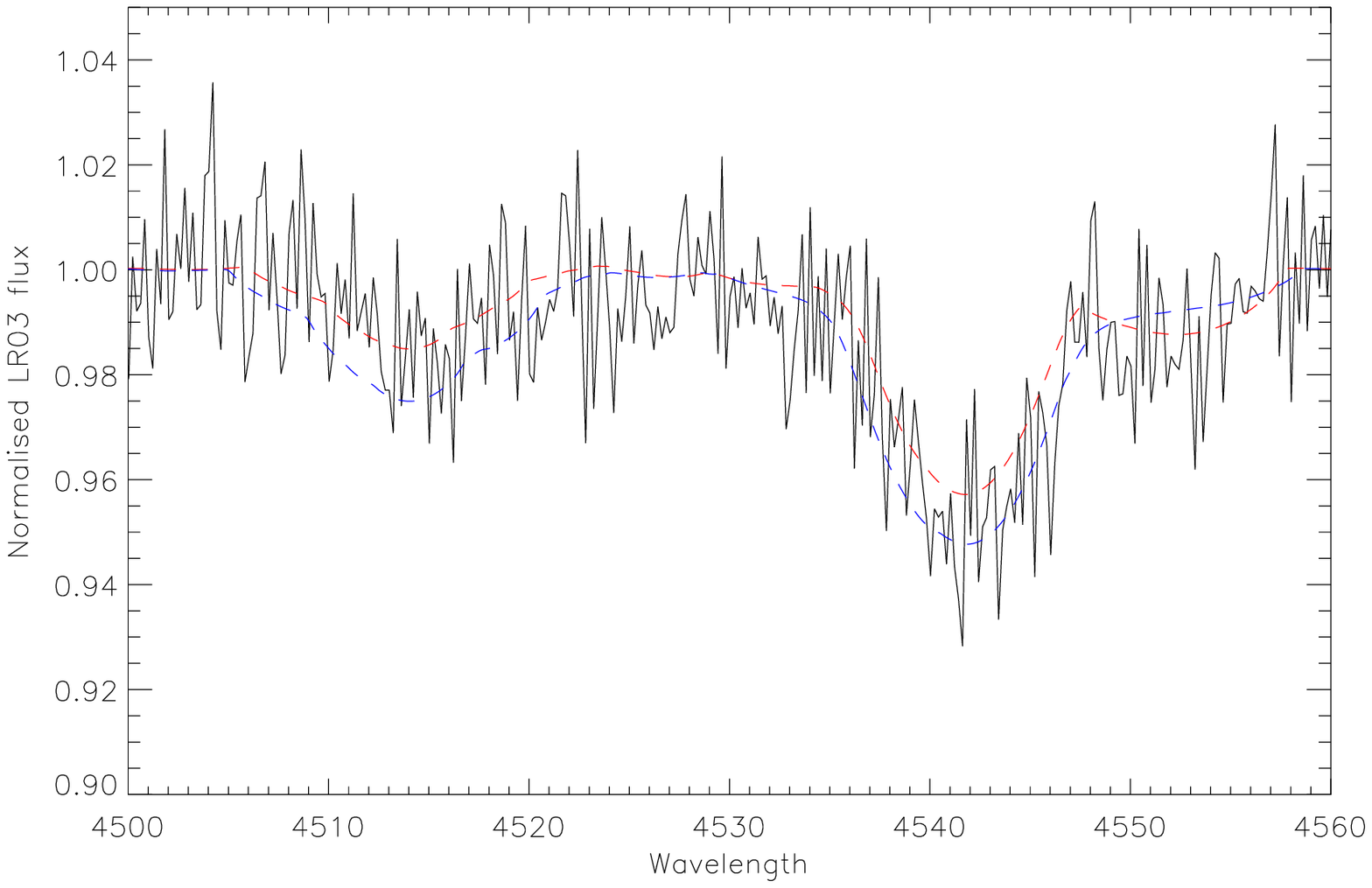}
\end{center}
\caption{LR02 (upper) and LR03 (lower) spectra  for VFTS\,399 encompassing the \ion{N}{iii} multiplet near 4515\AA. Also shown are theoretical spectra with nitrogen abundances 
of 7.62 dex (red, dash) and 7.92 (blue, dot-dash), convolved with an 
appropriate broadening function. The absorption near 4541\AA \ is due to 
\ion{He}{ii}. } \label{399_niii}
\end{figure}

\begin{table}
\caption{Bonnsai predictions for the current mass and lower limit for the age of VFTS\,399}
\begin{center}
\begin{tabular}{ccc}
\hline
\hline
log$(L_{\rm bol}/L_{\odot})$   & $M_{\rm curr}$   & Age   \\
                               & $(M_{\odot})$    & (Myr) \\
\hline
---         & $21.4_{-3.7}^{+2.7}$ & $>5.8_{-0.7}^{+0.6}$ \\ 
$4.7\pm0.1$ & $18.0_{-0.9}^{+1.2}$ & $>6.1_{-0.6}^{+0.5}$ \\
$4.9\pm0.1$ & $20.0_{-1.2}^{+1.6}$ & $>6.0_{-0.5}^{+0.4}$ \\
$5.2\pm0.1$ & $24.0_{-1.7}^{+2.2}$ & $>5.3_{-0.5}^{+0.5}$ \\
\hline
\end{tabular}
\end{center}
{All four models utilise stellar temperature, surface gravity and projected equatorial rotational velocity as input parameters, while rows 2-4 present the results of including the stellar luminosity estimates (column 1) as an additional constraint.  The quoted errors are $1\sigma$ uncertainties.}
\end{table}

\subsection{Understanding the X-ray emission}

Single OB stars are known to be (thermal) X-ray emitters, likely due 
to shocks propagating through their winds. However, this emission scales as log$(L_{\rm x}/L_{\rm bol})\sim10^{-7}$ (e.g. 
Berghoefer et al. \cite{berghoefer}, Sana et al. \cite{sana06}), while 
VFTS\,399  is orders of magnitude brighter, with log$(L_{\rm x}/L_{\rm bol})\sim0.3-4\times10^{-4}$ in the  
observations reported above, while the emission is of a non-thermal 
nature.  
These properties also exclude membership of the class of X-ray bright and variable $\gamma$ Cas-like OeBe stars, for which a high temperature thermal spectrum and log$(L_{\rm x}/L_{\rm bol})\sim 10^{-6}$ are expected
(Smith et al. \cite{smith}, Rauw et al. \cite{rauw}). Likewise, the peak X-ray luminosity of VFTS\,399  is  two orders of magnitude greater than 
that found for  Of?p stars and the subset of strongly  magnetic O-type 
stars which, in turn, are  X-ray 
over-luminous with respect to the wider population of massive stars (cf. Naz\'{e} et al. \cite{naze08}, Clark et 
al. \cite{clark09} and refs. therein). Additionally, 
the X-ray emission in such stars is once again thermal in origin, in contrast to VFTS\,399.

At its peak luminosity,  VFTS\,399  is an order of magnitude brighter than any O+O colliding wind binary, although seven binaries 
with luminous blue variable or Wolf-Rayet primaries demonstrate log$L_{\rm x}\geq10^{35}$erg~s$^{-1}$  
(Gagn\'{e} et al. \cite{gagne})\footnote{WR48a, Mk34, $\eta$ Car, R140a, WR25, R136c and CXO J1745-28.}. However in each of these 
systems the  primaries support significantly more powerful winds than 
are expected for VFTS\,399, suggesting that its X-ray flux is 
unlikely to arise via wind collisions (indeed  no signature of any  secondary is visible in our optical spectroscopic observations). 
Moreover the  X-ray emission from such systems is thermal in origin, with $kT\sim0.9-4.4$keV for those massive binaries with 
comparable X-ray fluxes. 

We are then left with the possibility that VFTS\,399  contains an accreting relativistic object and, given the nature of the primary, an obvious identification is with a Be X-ray binary. These systems are overwhelmingly comprised of  OeBe stars ($\sim$O9-B2 V-III; Negueruela \& Coe \cite{negueruela02}, McBride et al. \cite{mcbride})  orbited  by neutron stars and have 
periods $20$ days $\lesssim P_{\rm orb} \lesssim 400$ days and eccentricities $0.1 \lesssim e \lesssim0.9$ (Cheng et al. \cite{cheng}). The  interplay between orbital phase and the intrinsically variable circumstellar disc leads to complex, variable X-ray emission (e.g. Okazaki \& Negueruela \cite{okazaki})\footnote{Most Be X-ray binaries are transient systems, with outbursts typically $>10\times$  their quiescent X-ray luminosities.  Following the nomenclature of Stella et al. (\cite{stella}), Type I outbursts occur periodically during periastron passage and typically yield fluxes of $L_{\rm X}\sim 10^{36}-10^{37}$erg s$^{-1}$. Type II 
outbursts are much rarer, aperiodic, have a longer duration and reach  $L_{\rm X}\gtrsim 10^{37}$erg s$^{-1}$. 
In contrast, a subset of systems  with wide ($P_{\rm orb} \gtrsim 100$ days), low-eccentricity orbits
are  persistent sources with moderate luminosity ($L_{\rm X}\sim 10^{34}-10^{35}$erg s$^{-1}$)   and 
comparatively low-level variability ($\lesssim 10\times$ quiescent flux), of which X Per is an exemplar (Haberl 
et al. 
\cite{haberl98}, Reig \& Roche \cite{reig99}).}. Haberl et al. (\cite{haberl}) determined the  X-ray properties of a representative sample of SMC systems from single-epoch observations, finding the majority of spectra to be  best fit by  power laws
($\Gamma \sim 0.93_{-0.22}^{+0.34}$) with a wide range of absorbing columns ($\sim 10^{20} - 10^{22}$ cm$^{-2}$) and luminosities spanning $\sim 1.5\times10^{35} - 5.5\times10^{36}$ergs$^{-1}$. The X-ray properties of VFTS\,399  therefore appear entirely consistent with its classification as a Be X-ray binary.

 In such a scenario a  neutron star accretor would be strongly favoured on both theoretical and observational grounds. Only 
 one Be+black hole  X-ray binary has been identified from a Galactic population of $\sim$80 such systems 
 (Casares et al. \cite{casares}), while the same study shows it to have  a quiescent X-ray 
 luminosity some three orders of magnitude lower than VFTS\,399.

Mindful of the {\em caveats} in Sect. 2.2.1, the detection of  periodic  modulation of the X-ray lightcurve provides  support for such a conclusion, being most naturally explained as the pulsational period of a
rotating, accreting neutron star.  Importantly, both `short' ($\sim3.245$s) and `long' ($\sim2567$s) potential periodicities lie within the range of pulsational periods exhibited by neutron stars within  Be X-ray binaries (e.g. Knigge et al. \cite{knigge}, Reig et al. \cite{reig09}) and we expand upon this further in Sect. 4.1.

\section{VFTS\,399  in context}

An immediate question to ask is whether we may constrain the orbital parameters of the system from the extant data?  Adopting $M_{\rm 
curr}\sim18.0-20.0M_{\odot}$ for the system primary (Sect. 3.1), the observed RV variability (Table 2 and Sect. 2.1.2) implies an unexpectedly short orbital period of $P_{\rm orb}\sim4$ days for a canonical $1.4M_{\odot}$ neutron star companion. For $P_{\rm orb}\sim 20$days, consistent with the commonly found lower bound  for Be X-ray binaries (Cheng et al. \cite{cheng}), we would infer a companion mass of 
$\sim2.5M_{\odot}$, with any longer period strongly favouring a black hole, apparently excluded by the detection of X-ray pulsations\footnote{For comparison Antoniadis et al. (\cite{antoniadis}) determine $\sim2.01\pm0.04 M_{\odot}$ for the pulsar PSR J0348+0432, while Clark et al. (\cite{clark02}) report $\sim2.44\pm0.27 M_{\odot}$ for the non-pulsating relativistic accretor in 4U1700-37  and  Casares et al. (\cite{casares}) find
$\sim3.8-6.9 M_{\odot}$ for the black hole orbiting the Be star MWC656.}.  Preliminary 
Monte-Carlo analysis following the methodology of Sana et al. (\cite{sana13}) appears to favour a lower-mass companion with a short period. We caution that this finding should be 
regarded as provisional since it is  primarily driven by 
the discrepancy in RV between two epochs separated by  0.8days (Table 2; 2008 December 21, 22). Clearly further observations
in order to build an orbital RV curve for the system are required to address this issue. 

\subsection{The putative neutron star accretor}

Is the range of orbital periods suggested by the preceding analysis consistent with one or both of the potential pulsational periods? 
Following the relationship between the pulsational ($P_{\rm spin}$) and orbital ($P_{\rm orb}$) periods in Be X-ray 
 binaries delineated in the Corbet diagram (Corbet \cite{corbet}, Knigge et al. \cite{knigge}) we would infer $P_{orb}\lesssim20$~d for $P_{\rm spin}\sim3.245$~s and $P_{\rm orb}\sim500$~days for $P_{\rm spin}\sim2567$~s. While the first combination appears internally consistent, the long orbital period implied by the second would appear to be in tension with our RV data.

However, with $P_{orb}\lesssim19.25$~d and  $P_{\rm spin}\sim5560$s  (Reig et al. \cite{reig09}, \cite{reig12}), the O9.5 Ve+neutron star binary 4U 2206+54 demonstrates that such a combination of long pulsation but short orbital period is 
viable, although this would require VFTS\,399 to  host the third most slowly rotating neutron star of any  high mass X-ray binary, behind 4U 2206+54 and 2S 0114+650 ($P_{\rm orb}\sim11.59$day and $P_{\rm spin}\sim2.7$hr; Crampton et al. \cite{crampton}, Corbet et al. \cite{corbet99}, Farrell et al. \cite{farrell}). In this regard we note that both the X-ray luminosity and aperiodic variability of VFTS\,399 are replicated in 4U 2206+54 - which  is assumed to be powered by direct wind-fed accretion (Negueruela \& Reig \cite{negueruela}) -  while the primaries in both systems resemble one another 
(Blay et al. \cite{blay}, Walborn et al. \cite{walborn}).

In each of the systems named above the exceptionally long pulsational periods of the neutron stars are explicable under the assumption that  they had been 
significantly  spun down, first via magnetic dipole radiation and subsequently the propellor mechanism (Li \& van den Heuvel \cite{li}, Reig et al. \cite{reig09}, \cite{reig12}, Popov \& Turolla \cite{popov}). Operating on comparatively short timescales
($\sim10^4-10^5$yr; Popov \& Turolla \cite{popov}), this  would also be consistent with the relatively unevolved nature of 
VFTS\,399 (cf. footnote 6), although it would require extremely high magnetic fields, $\gtrsim10^{14}$G, in order to facilitate it (cf. Li \& van den Heuvel \cite{li} and Appendix A).

Nevertheless, under either scenario, it is  of interest that VFTS\,399 appears to host the second young pulsar within the 30 Dor complex after the
 isolated neutron star PSR J0537-6910 ($\sim$5arcmin/$\sim70$pc distant 
in projection). 
With $P_{\rm spin}\sim16$ms (Marshall et al. \cite{marshall}), PSR J0537-6910  rotates much more rapidly than the putative pulsar within VFTS\,399, naturally explicable because of  the lack of accretion driven spin-down in the former. However, if the VFTS\,399 pulsar is exceptionally  slowly rotating due to the effects of an extreme $B-$field one would be forced to explain their different field strengths ($B\gtrsim10^{14}$G versus $B\sim10^{12}$G for PSR J0537-6910; Marshall et al. \cite{marshall}), despite the likelihood of both pulsars  having formed from  the same underlying stellar population (cf. Dufton et al. \cite{dufton}).

\subsection{The O-type star mass donor}

VFTS\,399 demonstrates notable physical similarities with VFTS\,102, 
another member of the high rotational-velocity O-type star cohort of 30 
Dor (Ram\'{i}rez-Agudelo  et al. 
\cite{oscar}). Both are emission-line stars, share the same  spectral classification and 
 their temperature, luminosity and  surface gravity are identical within the errors quoted (Sect. 3.1 and Table 1 of Dufton et 
al. \cite{dufton}). Based on the proximity of VFTS\,102 to the isolated pulsar PSR J0537-6910 (Marshall et al. \cite{marshall}), Dufton et al. (\cite{dufton}) explored a common binary evolution scenario for both objects, with the $16M_{\odot}+15M_{\odot}$ 
 mass-transfer model of Cantiello et al. (\cite{cantiello}) yielding a rapidly rotating secondary well matched to VFTS\,102 at the point of supernova.

While it is tempting to adopt a comparable  pre-SN evolutionary scenario for VFTS\,399 -  albeit with the binary remaining 
bound at SN  - this would be hard to reconcile with the extant RV data.
Specifically, a  short orbital period would require the binary system to lose a 
large fraction of the mass and angular momentum released in the pre-SN mass-transfer process
  (Petrovic et al. \cite{petrovic}), since
 conservative evolution leads to orbital periods 
in the range $\sim50 - 300$~day (Wellstein et al. 2001).
In other words, this observational constraint implies that VFTS399 accreted comparatively little material, noting that 
the  accretion of only $\sim 2\,$M$_{\odot}$ is sufficient to spin it up to critical 
rotation. Had the initial mass ratio been close to one - as inferred for VFTS\,102 -  then the mass transfer 
would be expected to remain close to conservative (Langer \cite{langer}, de Mink et al. \cite{demink}). 
Therefore the properties of the VFTS399 binary system are best understood 
if the initial mass of the NS progenitor was significantly higher than the current mass of the Oe component. 

Given the current absence of a full orbital solution for VFTS\,399 the calculation of a tailored evolutionary model is 
premature, although based on previous analysis  we might expect an initial mass of $\gtrsim25M_{\odot}$ for the NS progenitor; 
 as a consequence of this we may speculate that VFTS\,399  might host a high-mass NS. 
Under the  mass-transfer scenario we favour the mass gainer is expected to quickly rejuvenate, and then behave just
as a single star of its new mass, with the possible exception of rapid rotation and surface abundance anomalies.
Therefore,  we might expect VFTS\,399 to have been born with broadly similar properties 
to those we infer for it now (Table 5 and Sect. 3.1), while our 
current age estimate  ($\sim6$Myr; Table 5) would be only  slightly lower than expected. Our hypothetical $\sim25M_{\odot}$ primary would
be expected to explode after $\sim7$Myr (Brott et al. \cite{brott}); fully consistent with such a scenario and  implying
 a rather recent ($\lesssim1$Myr) SN.

 As with VFTS\,102, we might therefore expect the systemic velocity of VFTS\,399 to have been influenced by the SN. Somewhat surprisingly, 
 there is  no indication of it being a radial velocity runaway, with  $v_{\rm rad} \sim 289.4\pm18.6$kms$^{-1}$ in comparison to a cluster 
 mean of $\sim 271$km$^{-1}$  and  a velocity dispersion $\sigma \sim10$kms$^{-1}$  (Sana et al. \cite{sana12}, \cite{sana13}). Nevertheless, 
 the location of VFTS\,399   - on the periphery of the complex and displaced from any recognised cluster (Fig. 1; Walborn et al. 
 \cite{walborn})  - suggests a runaway nature; our prediction of a recent SN would suggest a displacement of $\lesssim50$pc from its birth 
 site.  Ongoing  {\em Hubble Space Telescope} proper motions studies (PI: D.~J.~Lennon; see Sabbi et al. \cite{sabbi}) should provide a 
 complementary determination of the transverse velocity in order to test this prediction. 

Both VFTS\,102 and VFTS\,399 have been associated with a  population of 
apparently single, rapidly-rotating O-type  star runaways within the 30 
Dor complex 
(Walborn et al. \cite{walborn}).  Of these, 15 stars are of spectral type O8-O9.5 III-V\footnote{VFTS\,5 (O8 
Vn),   74 (O9 Vn), 91  (O9.5 IIIn), 102 (O9: Vnnne+), 138 (O9 Vn), 249 (O8 Vn), 399 (O9 IIIn), 530 (O9.5III:nn), 
531(O9.5III:nn), 574  (O9.5 IIIn), 592 (O9.5 Vn), 654 (O9 Vnn), 660 (O9.5 Vnn), 768 (O8 Vn) and  843 
(O9.5IIIn).} and hence might be  expected to exhibit the OeBe phenomenon (Negueruela  et al. \cite{iggy04}). 
However, at the time of our spectral  observations, only VFTS 102 and 399 were found to support the gaseous 
circumstellar discs from which a putative neutron star/black hole might accrete. Therefore  the 
remaining stars 
from this grouping could also  harbour relativistic companions which would remain undetectable at X-ray energies until they too formed circumstellar discs. Eldridge et al. (\cite{eldridge}) examined the pre- and post-SN evolution of massive binaries and concluded that they remain bound in $\sim20$\% of cases, with a most likely systemic velocity of $\sim50$ kms$^{-1}$. Given this prediction,  long term optical and X-ray observations of this cohort would appear warranted in order to determine if a subset of these stars are
 indeed quiescent X-ray binaries and hence received their anomalous rotational and/or radial velocities via binary interaction and subsequent 
SN kick.

\section{Concluding remarks}

The  multi-wavelength properties of the outlying, rapidly rotating O-type 
star 
VFTS\,399   appear difficult to reconcile with those expected for a single star, instead favouring a 
binary hypothesis. The X-ray properties strongly favour 
the presence of a relativistic  accretor, with the optical and near-IR 
properties consistent with classification  as a Be X-ray binary,  being the seventeenth 
such system  within the  LMC at the 
time of writing (Negueruela \& Coe \cite{negueruela02}, Vasilopoulos et al. \cite{vasil}) and the first 
identified within 30 Dor.

 The  discovery of an apparently robust, statistically-significant  periodicity in the X-ray lightcurve  provides strong evidence for a neutron star in the the system.
Unfortunately we are currently unable to  determine {\em conclusively} whether the $\sim2567$~s period reflects $P_{\rm spin}$ or instead is an alias of the true $P_{\rm spin}\sim3.245$~s, although the robustness of the former period to the secular evolution common in X-ray binaries argues that it is astrophysical in origin.
Breaking this degeneracy via observations with an alternative observatory such as {\em XMM} would be of considerable interest, since the former scenario would imply that the neutron star 
supported a magnetic field of comparable strength to those exhibited by magnetars in order to permit the dramatic spin down required to yield its current $P_{\rm spin}$.

These findings have a number of ramifications:\newline
(i) Identification of individual  (high-mass) 
X-ray binaries with their natal stellar aggregates is surprisingly  rare - 
 LS I +63 235 lies within the halo of the cluster NGC663,   
 while PSR  B1259-63 and IGR J00370+6122 are likely members of the Cen OB1 
and Cas OB8 associations respectively (Negueruela et al. \cite{iggy11}, 
Gonz\'{a}lez-Gal\'{a}n et al. \cite{gonzalez}). The location of VFTS\,399  within the confines of the 30 Dor complex could 
therefore aid
 the observational assessment of the predictions of current population-synthesis models. These include the production 
(and retention) rates of high mass X-ray binaries as a function of star-formation rate and the
duration of such activity. This would also inform studies of e.g.   the global X-ray luminosity/star-formation rate relationship for  non-active galaxies (e.g. Mineo et al. \cite{mineo}).\newline
 (ii) In conjunction with determinations of both radial and transverse velocity components, characterisation of the orbital period and eccentricity  of VFTS\,399 should permit the dynamics of the SN explosion to be reconstructed. Given that Pfahl et al. (\cite{pfahl}) and Podsiadlowski et al. (\cite{podsiadlowski}) suggest that the magnitude of kick is dependent on the pre-SN binary evolution,  the fact that we may place observational constraints on this for VFTS\,399  via comparison to the stellar population(s) of 30 Dor is of considerable interest.\newline
(iii) If the inference of an extreme $B-$field for the putative neutron star is confirmed, reconstruction of the pre-SN evolutionary pathway and SN dynamics will help constrain the formation requirements for such objects. Moreover, it would support the hypothesis that binarity is a common feature in the formation of magnetars, which in turn would arise from progenitors with a wide mass range 
from about $17M_{\odot}$ up to $40M_{\odot}$ or more (Sect. 4.2 and  Clark et al. \cite{clark14}). \newline 
(iv)  After VFTS102, VFTS\,399  is the second object drawn  from the 
population of rapidly rotating  apparently single O-type stars for which 
historical/current binarity has subsequently been inferred. Such  a finding is consistent with the hypothesis that binary interactions may play a key role in the formation of this cohort (de Mink et al. \cite{demink}, Ram\'{i}rez-Agudelo et al. \cite{oscar}).  If confirmed for further examples it will have considerable import for the production of the rapidly-rotating, chemically-homogeneous stars inferred to be the  progenitors of long gamma-ray bursts.

\begin{acknowledgements}
We thank the referee and  Danny Lennon and Ignacio Negueruela for their insightful comments which have greatly improved the paper. 
SdM acknowledges support by NASA through and Einstein Fellowship grant, PF3-140105.
 STScI is operated by AURA, Inc. under NASA contract NASA 5-26555. LKT and PSB were supported
 by Chandra X-ray Observatory general
observer grants GO4-15131X and GO5-6080X and by the Penn State ACIS
Instrument Team Contract SV4-74018, issued by the Chandra X-ray
Center, which is operated by the Smithsonian Astrophysical
Observatory for and on behalf of NASA under contract NAS8-03060. The
Guaranteed Time Observations included here were selected by the ACIS
Instrument Principal Investigator, Gordon P. Garmire, of the
Huntingdon Institute for X-ray Astronomy, LLC, which is under
contract to the Smithsonian Astrophysical Observatory; Contract
SV2-82024.

\end{acknowledgements}

{}

\appendix

\section{Spin-down in the presence of extreme magnetic fields}

Following the analysis of Li \& van den Heuvel (\cite{li}), the evolution of the spin 
period of the putative  neutron star companion to   VFTS\,399  depends on the accretion rate/torque  - a function of the (currently unknown) orbital separation and wind properties of the primary - {\em and} its magnetic field. For this mechanism to yield  $P_{\rm spin}\sim2567$~s it  must have been born with an (external)  magnetic field of comparable strength to those exhibited by  magnetars ($B\gtrsim 10^{14}$G). If we make the assumption that the neutron star in VFTS\,399 is rotating at its 
equilibrium period we may estimate the dipolar field strength from:\newline

$P_{\rm spin} \simeq (20 {\rm s})B_{12}^{6/7}\dot{M}_{15}^{-3/7}R_6^{18/7}M_{1.4}^{-5/7}$\newline

 (Li \& van den Heuvel \cite{li}, Reig et al. \cite{reig09}) where $B=10^{12}B_{12}$G is the dipolar field strength, $\dot{M}=10^{15}\dot{M}_{15}$g s$^{-1}$ the mass accretion rate and  $R=10^6R_6$cm and $M=1.4M_{1.4}$M$_{\odot}$ the neutron star radius and mass. We adopt $P_{\rm spin}\sim2567$~s, $R_6=M_{1.4}=1$ and $\dot{M}_{15}\approx L_XR/GM =0.25$ for $L_x \sim 0.5\times10^{35}$erg s$^{-1}$ (Reig et al. \cite{reig09}; Sect. 2.2), which leads to an order-of-magnitude estimate of $B\sim10^{14}$G for the neutron star. One might assume that a significantly lower accretion rate may alleviate the requirement for such a $B-$field. However, following Li and van den Heuvel (\cite{li})
e.g. $\dot{M}=10^{12}$g s$^{-1}$ (leading to $B=10^{13}$G) results in a spin-down timescale comparable to the total lifetime of VFTS\,399 (cf Dufton et al. \cite{dufton}), in contrast to its current relatively unevolved nature.

Recently, Shakura et al. (\cite{shakura}) proposed a new theoretical model for quasi-spherical accession onto slowly rotating 
neutron stars (see also Popov \& Turolla \cite{popov}). Reig et al. (\cite{reig12}) demonstrate that the accretor in 4U2206+54 still supports an extreme ($\sim10^{14}$G) $B-$field under this formulation, although this requires an anomalously slow primary wind (which is observed in this system; Rib\'{o} et al. \cite{ribo}).
Clearly, an orbital solution for VFTS\,399 and a determination of the wind parameters of the primary (mass loss 
rate and velocity at the orbital radius) in order to determine 
$\dot{M}$ would provide an empirical test of the extreme $B-$field hypothesis, as would a measure of the long term spin-period history of the putative neutron star, from which an estimate of $B$ may also be obtained.


\begin{thebibliography}{}

\bibitem[2013]{antoniadis}
Antoniadis, J., Freire, P. C. C., Wex, N. et al. 2013, Science, 340, 448


\bibitem[1996]{arnaud}
Arnaud, K. A. 1996, in Astronomical Society of the Pacific Conference Series,
Vol. 101, Astronomical Data Analysis Software and Systems V, ed. G. H.
Jacoby \& J. Barnes, 17

\bibitem[1997]{berghoefer}
Berghoefer, T. W., Schmitt, J. H. M. M., Danner, R. \& Cassinelli 1997, A\&A, 322, 167

\bibitem[2012]{bird}
Bird, A. J., Coe, M. J., McBride, V. A. \& Udalski, A. 2012, MNRAS, 423, 3663

\bibitem[1961]{blaauw}
Blaauw, A. 1961,  Bull. Astron. Inst. Netherlands, 15, 265

\bibitem[2006]{blay}
Blay, P., Negueruela, I., Reig, P. et al. 2006, A\&A, 446, 1095


\bibitem[2009]{bonanos}
Bonanos, A. Z., Massa, D. L., Sewilo, M., et al. 2009, AJ, 138, 1003

\bibitem[2011]{brott}
Brott, I., de Mink, S. E., Cantiello, M., et al. 2011, A\&A, 530, A115

\bibitem[2007]{cantiello}
Cantiello, M., Yoon, S.-C., Langer, N. \& Livio, M. 2007, A\&A, 465, L29

\bibitem[2014]{casares}
Casares, J., Negueruela, I., Rib\'{o}, M., et al. 2014, Nature, 505, 378


\bibitem[2014]{cheng}
Cheng, Z.-Q., Shao, Y. \& Li, X.-D. 2014, ApJ, 786, 128
\bibitem[2011]{cioni}
Cioni, M.-R. L., Clementini, G., Girardi, L., et al. 2011, A\&A, 527, A116

\bibitem[1999]{clark99}
Clark, J. S., Lyuty, V. M., Zaitseva, G. V. et al. 1999, MNRAS, 302, 167

\bibitem[2001]{clark01}
Clark, J. S., Tarasov, A. E., Okazaki, A. T., Roche, P. \&  Lyuty, V. M.   2001, A\&A, 380, 615

\bibitem[2002]{clark02}
Clark, J. S., Goodwin, S. P., Crowther, P. A. et al. 2002, A\&A, 392 909 

\bibitem[2003]{clark03}
Clark, J. S., Tarasov, A. E. \& Panko, E. A., 2003, A\&A, 403, 239

\bibitem[2009]{clark09}
Clark, J. S., Crowther, P. A. \& Mikles, V. J. 2009, A\&A, 507, 1567

\bibitem[2014]{clark14}
Clark, J. S., Ritchie, B. W., Najarro, F., Langer, N. \& Negueruela, I. 2014, A\&A, 565, A90

\bibitem[1986]{corbet}
Corbet, R. H. D. 1986, MNRAS, 220, 1047

\bibitem[1999]{corbet99}
Corbet, R. H. D., Finley, J. P. \& Peele, A. G. 1999, ApJ, 511, 876

\bibitem[2005]{corcoran}
Corcoran, M. F. 2005, AJ, 129, 2018

\bibitem[1985]{crampton}
Crampton, D., Hutchings, J. B. \& Cowley, A. P. 1985, ApJ, 299, 839

\bibitem[2007]{demink07}
de Mink, S. E., Pols, O. R. \& Hilditch, R. W. 2007, A\&A, 467, 1181

\bibitem[2013]{demink}
de Mink, S. E., Langer, N., Izzard, R. G., Sana, H. \& de Koter, A. 
2013, ApJ, 764, 166


\bibitem[1994]{dougherty}
Dougherty, S. M., Waters, L. B. F. M., Burki, G., et al. 1994, A\&A, 290, 609

\bibitem[2011]{dufton}
Dufton, P. L., Dunstall, P. R., Evans, C. J. et al. 2011, ApJ, 743, L22 

\bibitem[2013]{dufton13}
Dufton, P. L., Langer, N., Dunstall, P. R., et al. 2013, A\&A, 550, A109

\bibitem[2011]{eldridge}
Eldridge, J. J., Langer, N. \& Tout, C. A. 2011, MNRAS, 414, 3501

\bibitem[2011]{evans}
Evans, C. J., Taylor, W. D., H\'{e}nault-Brunet, V., et al. 2011, A\&A, 530, A108

\bibitem[2008]{farrell}
Farrell, S. A., Sood, R. K., O'Neill, P. M. \& Dieters, S. 2008, MNRAS, 389, 608



\bibitem[2012]{gagne}
Gagn\'{e}, M.,  Fehon, G. Savoy, M. R. et al. 2012, in Proceedings of Four Decades of Massive Star Research - A Scientific Meeting in Honour of Anthony F. J. Moffat, ASP 
Conference Series, Vol. 465. San Francisco: Astronomical Society of the Pacific, 2012., p.301

\bibitem[2008]{galache}
Galache, J. L., Corbet, R. H. D., Coe, M. J. et al. 2008, ApJS, 177, 189




\bibitem[2000]{gibson}
Gibson, B. K., 2000, Mem. Soc. Astron. Ital., 71, 693


\bibitem[2014]{gonzalez}
Gonz\'{a}lez-Gal\'{a}n, A., Negueruela, I., Castro, N., et al. 2014, A\&A, 
556, A131

\bibitem[2000]{grebel}
Grebel, E. K. \& Chu, Y.-H. 2000, AJ, 119, 787



\bibitem[1998]{haberl98}
Haberl, F., Angelini, L., Motch, C. \& White, N. E. 1998, A\&A, 330, 189

\bibitem[2008]{haberl}
Haberl, F., Eger, P. \& Pietsch, W. 2008, A\&A, 489, 327

\bibitem[2002]{hurley}
Hurley, J. R., Tout, C. A. \& Pols, O. R. 2002, MNRAS, 329, 897

\bibitem[2007]{kato}
Kato, D., Nagashima, C., Nagayama, T., et al. 2007, PASJ, 59, 615



\bibitem[2011]{knigge}
Knigge, C., Coe, M. J. \& Podsiadlowski, P. 2011, Nature, 479, 372

\bibitem[2014]{koehler}
K\"{o}hler, K., Langer, N., de Koter, A., et al. 2015, A\&A, 573, A71

\bibitem[2012]{langer}
Langer, N. 2012, ARAA, 50, 107

\bibitem[2003]{lanz}
Lanz, T. \& Hubeny, I., 2003, ApJS, 146, 417

\bibitem[1999]{li}
Li, X.-D. \& van den Heuvel, E. P. J. 1999, ApJ, 513, L45

\bibitem[1976]{Lomb76}
Lomb N.~R., 1976, Ap\&SS, 39, 447 



\bibitem[2004]{maiz04}
Ma\'{i}z Apell\'{a}niz, J. 2004, PASP, 116, 859

\bibitem[2014]{maiz14}
Ma\'{i}z Apell\'{a}niz, J., Evans, C. J., Barb\'{a}, R. H. et al. 2014, A\&A, 564, A63

\bibitem[1998]{marshall}
Marshall, F. E., Gotthelf, E. V., Zhang, W., Middleditch, J. \& Wang Q. D.
1998, ApJ, 499, L179

\bibitem[2005]{martins}
Martins, F., Schaerer, D. \& Hillier, D. J. 2005, A\&A, 436, 1065

\bibitem[2006]{martins06}
Martins, F. \& Plez, B. 2006, A\&A, 457, 637

\bibitem[2008]{mcbride}
McBride, V. A., Coe, M. J., Negueruela, I. Schurch, M. P. E.   \& McGowan, K. E. 2008, MNRAS, 388, 1198

\bibitem[2006]{meixner}
Meixner, M., Gordon, K. D., Indebetouw, R. et al. 2006, AJ, 132, 2268

\bibitem[2002]{mennickent}
Mennickent, R. E., Pietrzy\'{n}ski, G., Geiren, W. \& Szewczyk, O. 2002, A\&A, 393, 887

\bibitem[2012]{mineo}
Mineo, S., Gilfanov, M. \& Sunyaev, R. 2012, MNRAS, 419, 2095

\bibitem[2008]{naze08}
Naz\'{e}, Y., Walborn, N. R. \& Martins, F. 2008, RMxAA, 44, 331


\bibitem[2001]{negueruela}
Negueruela, I. \& Reig, P. 2001, A\&A, 371, 1056

\bibitem[2002]{negueruela02}
Negueruela, I. \& Coe, M. J. 2002, A\&A, 385, 517

\bibitem[2004]{iggy04}
Negueruela, I., Steele, I. A. \&  Bernabeu G. 2004, AN, 325, 749


\bibitem[2011]{iggy11}
Negueruela, I. Rib\'o, M., Herrero, A. et al. 2011, ApJ, 732, L11

\bibitem[2001]{okazaki}
Okazaki, A. T., \& Negueruela, I., 2001, A\&A, 377, 161

\bibitem[2005]{petrovic}
Petrovic, J., Langer, N. \& van der Hucht, K. A. 2005, A\&A, 435, 1013


\bibitem[2002]{pfahl}
Pfahl, E. Rappaport, S., Podsiadlowski, P. \& Spruit, H.
2002, ApJ, 574, 364

\bibitem[2004]{podsiadlowski}
Podsiadlowski, Ph., Langer, N., Poelerends, A. J. T., et al. 2004, ApJ, 612, 1044

\bibitem[2012]{popov}
Popov, S. B. \& Turolla, R. 2012, MNRAS, 421, L127

\bibitem[1989]{Press89}
Press W.~H. \&  Rybicki G.~B., 1989, ApJ, 338, 277 



\bibitem[2005]{puls}
Puls, J., Urbaneja, M. A., Venero, R. et al. 2005, A\&A, 435, 669

\bibitem[2013]{oscar}
Ram\'{i}rez-Agudelo, O. H., Sim\'{o}n-D\'{i}az, S., Sana, H., et al. 2013, A\&A, 560, A29

\bibitem[2013]{rauw}
Rauw, G., Naz\'{e}, Y., Spano, M., Morel, T., ud-Doula, A. 2013, A\&A, 555, L9

\bibitem[2007]{reig}
Reig, P. 2007, MNRAS, 377, 867

\bibitem[2007]{reig99}
Reig, P. \& Roche, P. 1999, MNRAS, 306, 100

\bibitem[2009]{reig09}
Reig, P., Torrej\'{o}n, J. M., Negueruela, I., et al. 2009, A\&A, 494, 1073

\bibitem[2012]{reig12}
Reig, P., Torrej\'{o}n, J. M. \& Blay, P. 2012, MNRAS, 425, 595 

\bibitem[2006]{ribo}
Rib\'{o}, M., Negueruela, I., Blay, P., Torrej\'{o}n, J. M. \& Reig, P. 2006,
A\&A, 449, 687

\bibitem[2012]{riquelme}
Riquelme, M. S., Torrej\'{o}n, J. M. \& Negueruela, I. 2012, A\&A, 539, A114

\bibitem[2013]{sabbi}
Sabbi, E., Anderson, J., Lennon, D. J., et al. 2013, AJ, 146, 53

\bibitem[1955]{salpeter}
Salpeter, E. E., 1955, ApJ, 121, 161

\bibitem[2006]{sana06}
Sana, H., Rauw, G., Naz\'{e}, Y., Goset, E. \& Vreux, J.-M.
2006, MNRAS, 372, 661

\bibitem[2012]{sana12}
Sana, H., Dunstall, P. R., H\'{e}nault-Brunet, V. et al. 2012, in Proceedings of Four Decades of Massive Star Research - A Scientific Meeting in Honour of Anthony F. J. Moffat, ASP 
Conference Series, Vol. 465. San Francisco: Astronomical Society of the Pacific, 2012., p.284

\bibitem[2013]{sana13}
Sana, H., de Koter, A., de Mink, S. E., et al. 2013, A\&A, 550, A107

\bibitem[1982]{Scargle82} 
Scargle J.~D., 1982, ApJ, 263, 835 


\bibitem[2014a]{rejuvenation}
Schneider, F. R. N., Izzard, R. G., de Mink, S. E., et al. 2014, ApJ, 780, 117

\bibitem[2014b]{schneider}
Schneider, F. R. N., Langer, N., de Koter, A.,  et al. 2014b, A\&A, 570, A66

\bibitem[2012]{shakura}
Shakura, N., Postnov, K., Kochetkova, A \& Hjalmarsdotter, L. 2012, MNRAS, 420, 216


\bibitem[2005]{shtyk}
Shtykovskiy, P. \& Gilfanov, M. 2005, A\&A, 431, 597

\bibitem[2005]{sidoli}
Sidoli, L., Mereghetti, S., Larsson, S. et al. 2005, A\&A, 440, 1033

\bibitem[2006]{skrutskie}
Skrutskie, M. F., Cutri, R. M., Stiening, R. et al. 2006, AJ, 131, 1163

\bibitem[2012]{smith}
Smith, M. A., Lopes de Oliveira, R., Motch, C., et al. 2012, A\&A, 540, A53

\bibitem[1986]{stella}
Stella, L., White, N. E. \& Rosner, R. 1986, ApJ, 308, 669

\bibitem[2013]{sturm}
Sturm, R., Haberl, F., Pietsch, W. et al. 2013, A\&A, 558, A3

\bibitem[1995]{tarasov}
Tarasov, A. E. \& Roche, P. 1995, MNRAS, 276, L19

\bibitem[2004]{townsend}
Townsend, R. H. D., Owocki, S. P. \& Howarth, I. D. 2004, MNRAS,  350, 189


\bibitem[2006]{townsley06}
Townsley, L. K., Broos, P. S., Feigelson, E. D., Garmire, G. P. \& Getman, K. V. 2006, AJ, 131, 2164 

\bibitem[2014]{townsley14}
Townsley, L. K., Broos, P. S., Garmire, G. P., et al. 2014, ApJS, 213, 1

\bibitem[2003]{udalski}
Udalski, A. 2003, AcA, 53, 291


\bibitem[2013]{vasil}
Vasilopoulos, G., Maggi, p., Haberl, F. et al. 2013, A\&A, 558, A3

\bibitem[1997]{walborn97}
Walborn, N. R. \& Blades, J. C. 1997, ApJS, 112, 457

\bibitem[2013]{walborn13}
Walborn, N. R., Barb\'{a}, R. H. \& Sewi{\l}o, M. M. 2013, AJ, 145, 98

\bibitem[2014]{walborn}
Walborn, N. R., Sana, H.,  Sim\'{o}n-D\'{i}az, S., et al. 2014, A\&A, 564, A40

\bibitem[2001]{wellstein}
Wellstein, S., Langer, N. \& Braun, H. 2001, A\&A, 369, 939





\end{thebibliography}
\end{document}